\documentstyle[aps,preprint,pra]{revtex}
\tighten

\begin{document}
\draft
\title{Higher order processes in electromagnetic production of
electron positron pairs in relativistic heavy ion collisions}
\author{
Kai Hencken \and
Dirk Trautmann}
\address{
Institut f\"ur theoretische Physik der Universit\"at Basel,
Klingelbergstrasse 82, 4056 Basel, Switzerland
}
\author{
Gerhard Baur
}
\address{
Institut f\"ur Kernphysik (Theorie), Forschungszentrum J\"ulich,
52425 J\"ulich, Germany
}
\date{\today}

\maketitle

\begin{abstract}
We study higher-order effects in the electromagnetic production of
electron-positron pairs in relativistic heavy ion collisions.
Treating the field of the heavy ions as an external field and
neglecting the interaction among electrons and positrons, we show
that the $N$-pair creation amplitude is the antisymmetrised product
of $N$ one-pair creation amplitudes and the vacuum amplitude.
Neglecting contributions coming from exchange terms, we show that
the total probability for $N$ pairs is approximately a Poisson
distribution.  We investigate further the structure of the reduced
one-pair amplitude, concentrating especially on multiple-particle
corrections.  We calculate the first of these corrections in second
order Magnus theory based on our previous result in second-order
Born approximation for impact parameter $b$ zero.  Explicit
calculations show that the total probability is increased up to 10
\% by this correction for realistic collider parameters.  The
calculations can also be used to confirm the use of the Poisson
distribution for the total probability.
\end{abstract}


\narrowtext

\section{Introduction}
The electromagnetic production of electron-positron pairs in
relativistic heavy ion collisions has gained some interest recently
due to the observation that the total probability violates unitarity
if calculated in lowest order even for realistic energies and impact
parameters as large as the Compton wavelength
\cite{Baur:Bertulani:88}.  It can be expected therefore that
higher-order effects --- especially the multiple-pair production ---
are of importance.  Several models have been used in order to cure
the unitarity violation and predict multiple-pair probabilities,
based on the summation of some subclasses of
diagrams\cite{Baur:Unitarity:90,Baur:letter:90,Rhoades-Brown:91,Best:92}.
 All authors find that the pair multiplicity can be approximated by
a Poisson distribution and that the lowest-order results should be
interpreted as the pair multiplicity.  Unitarity is no longer
violated due to the inclusion of the vacuum amplitude, which reduces
all $N$-pair probabilities by the same factor.  Recently Ionescu
\cite{Ionescu:93} has proposed a model, which differs from the
previous ones, but which is not applicable for multiple-pair
production.

We are going to extend these existing models into two directions:
First we address the question how the $N$-pair creation amplitude is
exactly reducible to one-pair creation amplitudes and what
approximations are necessary in order to get the Poisson
distribution.  We then compare with earlier results.  Secondly we
study higher-order processes contributing to the lowest-order
one-pair creation amplitude.  We concentrate on multiple-particle
effects, which have not been included in previous models.

This paper is arranged as follows: In Sec.~\ref{Feynman} we show
that the $N$-pair creation amplitude can be written as an
antisymmetrised product of reduced one-pair creation amplitudes and
the vacuum amplitude.  For this we use Feynman boundary conditions
for the fermion field.  Neglecting contributions from exchange terms
in the calculation of the total probability we get a Poisson
distribution for the total $N$-pair probability quite easily.  In
Sec.~\ref{Sec:Pertubation} the same result is derived using the
perturbation expansion of the $S$-matrix, which gives us some
insight into the processes contributing to the reduced pair creation
amplitude and the vacuum amplitude.  Finally we compare in
Sec.~\ref{Sec:Comparision} our results with other existing ones.

In Sec.~\ref{Sec:Magnus} we derive the general form of the
$S$-operator in second-order Magnus theory and use it in
Sec.~\ref{Sec:Highercorr} in order to extract from this the
lowest-order multiple-particle correction to the one-pair creation
probability.  This correction is then calculated in
Sec.~\ref{BornZero} based on our analytic form of the differential
pair creation amplitude in second-order Born approximation for
impact parameter $b$ zero\cite{BornZero}.  The same calculation is
used also to study the deviation of the total two-pair creation
probability from the Poisson distribution, which is found to be
rather small.  Results of these calculations together with
conclusions are then summarized in Sec.~\ref{Results}.

In the Appendix, we show a way to calculate the coefficients
appearing in the Magnus theory expansion, which can therefore be
calculated to arbitrary order.

Throughout this paper, we treat the fields of the heavy ions as
external fields.  We neglect the electromagnetic interaction between
electrons and positrons.  This seems to be justified for almost all
fermions, apart from electrons and positrons, which are produced
with very low relative velocitiy, and even in a bound $e^+$--$e^-$
state, due to the smallness of the coupling constant $\alpha$
between the fermions compared to the effective coupling constant
$Z\alpha$ for the interaction with the external field.  Due to these
assumptions, our many particle system is essentially a system of
independent fermions in an external field.  For such a system the
$S$ operator is known to be of the form of a time-ordered
exponential
\begin{eqnarray}
S &=& {\cal T} \exp\left\{ - i \int_{-\infty}^{+\infty} H_I(t) dt
  \right\} \nonumber\\
  &=& {\cal T} \exp\left\{ \int d^4x : \bar\Psi(x)
  [-i e \not\!\!A(x)] \Psi(x) : \right\}
\label{Intro:SMatrix}
,
\end{eqnarray}
where $H_I$ is the Hamiltonian in the interaction picture.  This
form of the $S$ operator is the starting point for our calculations.

The general theory of a fermion interacting with an external
electromagnetic field is rather old and has been developed already
{}from the beginning of QED.  It has been developed especially by
Feynman \cite{Feynman:49} and Schwinger \cite{Schwinger:54}.  An
overview can also be found in
\cite{Itzykson:Zuber:80,Bialynicki-Birula:75,Schweber:61}.

Although we discuss only the electromagnetic pair production by
heavy ion collisions, the general theory can also be applied to
other pair production processes in external fields, for example,
pair creation due to bremsstrahlung\cite{Lippert:91}.

\section{Reduction using Feynman boundary conditions}
\label{Feynman}

The amplitude for the $N$-pair creation is given with the help of
the $S$ operator above as
\begin{equation}
S_{fi} = \left< f \right| S \left| i \right>
{}.
\end{equation}
As we want to calculate pair creation, the initial state is the
vacuum state $\left| 0 \right>$ and the final state an $N$-pair
state given by
\begin{equation}
\left| f \right> = b^+_{k_1} d^+_{l_1}\cdots
  b^+_{k_N} d^+_{l_N} \left| 0 \right>
,
\end{equation}
where $k_i$ denotes the electron quantum numbers, for example,
momentum and spin, and $l_i$ the same for the positrons.  The $S$
matrix is therefore
\begin{equation}
S_N := S(k_1, l_1\cdots,k_N,l_N)
  = \left< 0 \right| d_{l_N} b_{k_N}
  \cdots d_{l_1} b_{k_1} S \left| 0 \right>
{}.
\label{Eq:Sfi}
\end{equation}

If we neglect the interaction of the electrons and positrons among
each other, we are essentially dealing with a system of independent
particles.  Therefore the field operator at any given time is
linearly connected to its value at the boundary.  As we can
decompose the field operator into creation and annihilation
operators, the same is also true for them.

Using retarded boundary conditions this means that the creation and
annihilation operators for electrons and positrons at $t\rightarrow
+ \infty$ are connected through a unitary matrix with those at
$t\rightarrow -\infty$.  In the interaction picture this
transformation is expressed as
\begin{mathletters}
\begin{eqnarray}
b_p S & = & \sum_{i>0} a_{pi}\  S b_i +
  \sum_{j<0} a_{pj}\  S d_j^+,\\
d_q S & = & \sum_{i>0} a_{qi}^*\  S b_i^+ +
  \sum_{j<0} a_{qj}^*\  S d_j
{}.
\end{eqnarray}
\end{mathletters}

The main idea is now not to use these boundary conditions but
Feynman boundary conditions instead.  These are of a mixed retarded
and advanced type.  With these the electron creation operators for
$t\rightarrow + \infty$ are connected with the electron creation
operators for $t\rightarrow -\infty$ and the positron annihilation
operators for $t\rightarrow + \infty$.  This corresponds to the
St\"uckelberg-Feynman interpretation of electrons moving forward,
positrons moving backwards in time.  Again in the interaction
picture this is written as
\cite{Bialynicki-Birula:75,Feynman:QED,Greiner:85}
\begin{equation}
b_p S = \sum_{q>0} s^{++}_{pq} S b_q +
  \sum_{q'<0} s^{+-}_{pq'} d_{q'}^+ S
\label{Feynman:bpS}
{}.
\end{equation}
It has been shown already by Feynman that these boundary conditions
are completely equivalent to the retarded ones, but that they are
better suited for QED \cite{Feynman:49}.  For this relation to be
applicable, we assume also that the external field vanishes
asymptotically.

Similar relations exist for all other creation and annihilation
operators also; Eq.~(\ref{Feynman:bpS}) is the only relation we will
need in the following.  Applying it to $b_{k_1}$ we get for $S_N$:
\begin{eqnarray}
S_N &=& \left< 0 \right| d_{l_N} b_{k_N}
  \cdots b_{k_2} d_{l_1} \nonumber\\
& &\times  \left[\sum_{q_1>0} s^{++}_{k_1q_1} S b_{q_1} +
  \sum_{q_1'<0} s^{+-}_{k_1q_1'} d_{q_1'}^+ S\right]
  \left| 0 \right>
{}.
\end{eqnarray}

The $b_{q_1}$ operator annihilates the vacuum state, therefore we
drop this term:
\begin{eqnarray}
S_N&=& \left< 0 \right| d_{l_N} b_{k_N}
  \cdots b_{k_2} d_{l_1}
        \sum_{q_1'<0} s^{+-}_{k_1q_1'} d_{q_1'}^+ S
        \left| 0 \right> \nonumber \\
&=& \sum_{q_1'<0} s^{+-}_{k_1q_1'}
  \left< 0 \right| d_{l_N} b_{k_N}
  \cdots b_{k_2} d_{l_1}
  d_{q_1'}^+ S \left| 0 \right>
{}.
\end{eqnarray}
Using the fact that $b_{k_2}$ anticommutes with $d_{l_1}$ and
$d^+_{q_1'}$ we get
\begin{equation}
S_N = \sum_{q_1'<0} s^{+-}_{k_1q_1'}
  \left< 0 \right| d_{l_N} b_{k_N} \cdots
  d_{l_1} d_{q_1'}^+ b_{k_2} S \left| 0 \right>
{}.
\end{equation}

Now one replaces $b_{k_2}$ using again Eq.~(\ref{Feynman:bpS}) and
then commutes the next $b_{k_i}$ to the most right.  Doing this for
all $b_{k_i}$ we finally get:
\begin{eqnarray}
S_N &=& \sum_{q_1'\cdots q_N'<0} s^{+-}_{k_1q_1'} \cdots
  s^{+-}_{k_Nq_N'} \nonumber\\
& &\times  \left< 0 \right| d_{l_N} \cdots
  d_{l_1} d^+_{q_1'} \cdots d^+_{q_N'} S
  \left| 0 \right>
{}.
\end{eqnarray}

The vacuum expectation value can be calculated easily using the
anticommutation relations
\begin{equation}
\left\{d_{l_i}, d^+_{q_j'}\right\} = \delta_{l_i,q_j'}
\end{equation}
in order to move $d^+_{q_1'}$ to the left, until it reaches the left
side, where it annihilates the vacuum.  Doing this for all creation
operators we see that we get a nonvanishing result only, if every
$q_i'$ is equal to some $l_j$, and the overall sign of the vacuum
expectation value is equal to the signum of the permutation to
arrange $l_1 \cdots l_N$ in the same sequence as the corresponding
$q_1' \cdots q_N'$.  We get
\begin{eqnarray}
\lefteqn{\left< 0 \right| d_{l_N} \cdots
  d_{l_1} d^+_{q_1'} \cdots d^+_{q_N'} S
  \left| 0 \right> =} \qquad \nonumber\\
&&  \left< 0 \right| S \left| 0 \right>
  \sum_{\sigma} \mbox{sgn}(\sigma)
  \delta(l_{\sigma(1)},q_1') \cdots \delta(l_{\sigma(N)},q_N')
\end{eqnarray}
and therefore for $S_N$:
\begin{eqnarray}
\lefteqn{S_N(k_1,l_1,\cdots,k_N,l_N) =} \qquad \nonumber\\
&&  \left< 0 \right| S \left| 0 \right>
  \sum_{\sigma} \mbox{sgn}(\sigma)
  s^{+-}_{k_1l_{\sigma(1)}} \cdots
  s^{+-}_{k_Nl_{\sigma(N)}}
,
\end{eqnarray}
where the sum consists of $N!$ terms.  This can also be written
formally as
\begin{equation}
S_N = \left< 0 \right| S \left| 0 \right>
\det\left[ s^{+-}_{k_i l_j}\right]
{}.
\end{equation}

This is the main result, which we will use.  The $N$-pair creation
amplitude can be written as an antisymmetrised product of the
$s^{+-}_{k l}$ times the vacuum amplitude.  It is easy to see that
the $s^{+-}_{k l}$ are just the one-pair creation amplitudes divided
by the vacuum amplitude.  In the following we call this amplitude
the ``reduced'' amplitude.

Normally the vacuum amplitude does not show up in calculations, as
it is often of magnitude one, for example, for time independent
systems.  In our case the external fields are explicitly time
dependent, therefore the vacuum amplitude can not be neglected.
This can be understood easily, if one notices that the vacuum
amplitude is the amplitude for the ``no pair creation process'',
that is, the probability amplitude for the vacuum to remain the
vacuum.  As pair creation does occur, this probability must be
smaller than one and therefore the amplitude has to be of magnitude
less than one.  Also we see here that the vacuum amplitude appears
as a factor in all $N$-pair amplitudes and therefore reduces all
these amplitudes by the same factor.

{}From this we can get easily an approximate Poisson distribution for
the total $N$-pair probability, which has also been found by earlier
calculations.  For this we need the absolute value squared of $S_N$.
As $S_N$ consists of $N!$ terms, we get a total of
$\left(N!\right)^2$ terms.  In most of these terms we have the
product coming from two different permutations.  Normally the
produced electrons and positrons are correlated to some extent with
each other.  If we therefore assume that terms, where two different
permutations appear, are much smaller than those with the same
permutation and can therefore be neglected, we get only $N!$ terms:
\begin{equation}
P(N) \approx \left| \left< 0 \right| S \left| 0 \right> \right|^2
  \sum_{\sigma} \left| s^{+-}_{k_1l_{\sigma(1)}}\right|^2
  \cdots
  \left| s^{+-}_{k_Nl_{\sigma(N)}}\right|^2
{}.
\end{equation}
Summing or integrating now over all states we get for the total
probability
\begin{eqnarray}
\lefteqn{P_{\rm total}(N) \approx \frac{1}{\left(N!\right)^2}
  \left| \left< 0 \right| S \left| 0 \right> \right|^2}
  \qquad \nonumber\\
&&\times  \sum_{k_1,l_1,\cdots,k_N,l_N}
  \sum_{\sigma} \left| s^{+-}_{k_1l_{\sigma(1)}}\right|^2
  \cdots
  \left| s^{+-}_{k_Nl_{\sigma(N)}}\right|^2
,
\end{eqnarray}
where the factor $1/\left(N!\right)^2$ has been introduced in order
to correct for the multiple summation over electrons and positrons.
As every term in the sum over $\sigma$ gives the same result, we get
\begin{eqnarray}
P_{\rm total}(N)&\approx&\frac{1}{N!} \left|
  \left< 0 \right| S \left| 0 \right> \right|^2
  \sum_{k_1,l_1} \left| s^{+-}_{k_1l_1}\right|^2
  \cdots
  \sum_{k_N,l_N} \left| s^{+-}_{k_Nl_N}\right|^2
\nonumber\\
&=& \frac{1}{N!} \left| \left< 0 \right| S \left| 0 \right> \right|^2
  \left[\sum_{k,l} \left| s^{+-}_{kl}\right|^2\right]^N
\nonumber\\
&=& P(0) \frac{\left[ P_{\rm total}^R \right]^N}{N!}
,
\end{eqnarray}
with $P(0)$ the ``no pair'' probability and $P_{\rm total}^R =
\sum_{k,l} \left| s^{+-}_{kl} \right|^2$ the total reduced one-pair
probability.  This is just a Poisson distribution, and $P(0)$ can be
calculated using the fact that the sum over all $N$ has to be one:
\begin{equation}
P(0) \sum_N \frac{\left[ P_{\rm total}^R \right]^N}{N!}
= P(0) \exp\left[P_{\rm total}^R\right] = 1
{}.
\end{equation}
{}From this we finally get
\begin{equation}
P_{\rm total}(N) = \exp\left[- P_{\rm total}^R\right]
\frac{\left[ P_{\rm total}^R \right]^N}{N!}
{}.
\end{equation}

The only approximation that was necessary in order to get this
result was the neglection of all ``exchange terms'' in the
calculation of the probability.  We show later that at least for the
two-pair creation this neglection seems to be justified.

Therefore for the calculation of the $N$-pair creation probability
it suffices to calculate the reduced one-pair creation probability.
All other probabilities are given then by the Poisson distribution.
If one wants to go beyond the Poisson distribution, it suffices to
calculate the reduced one-pair creation amplitude and the vacuum
amplitude.

This derivation is rather simple, but it is not useful for explicit
calculations.  Especially the nature of the processes contributing
to the reduced pair creation amplitude as well as the vacuum
amplitude remains unexplained.  Therefore we derive the same result
in the next section using the perturbation-theory expansion of the
$S$ operator together with the Wick theorem.

\section{Reduction using perturbation theory}
\label{Sec:Pertubation}

In this section we use the field operator $\Psi(x)$ directly instead
of the creation and annihilation operators as in the previous
section, that is, we are going to calculate
\begin{equation}
S_N(y_1,\cdots,y_N,y_1',\cdots,y_N')
  = \left< f \right| S \left| i \right>
,
\end{equation}
where the initial state is again the vacuum $\left| 0 \right>$, and
the final state is
\begin{equation}
\left| f \right> = \bar\Psi^{(+)}(y_N) \Psi^{(-)}(y_N') \cdots
  \bar\Psi^{(+)}(y_1) \Psi^{(-)}(y_1') \left| 0 \right>
{}.
\end{equation}
Here we have made use of the frequency parts of the field operators
for electrons and positrons.  The electron coordinates are $y_i$,
the positron coordinates $y'_i$ and the time coordinate is the same
for all coordinates and assumed to go to $\infty$.  $S_N$ gives the
probability amplitude in occupation number space to get electrons
finally at $y_i$ and positrons at $y_i'$.  We rewrite the final
state a little bit by introducing normal ordering of the field
operators:
\begin{equation}
\left< f \right| = \left< 0 \right|
: \bar\Psi(y'_1) \Psi(y_1) \cdots \bar \Psi(y'_N) \Psi(y_N) :
\ ,
\end{equation}
where we can use the whole field operators instead of the frequency
parts, as the normal ordering together with the vacuum state
guarantees that only the appropriate part of each field operator
contributes.  Throughout the calculation in this section the Dirac
indices will be suppressed.  In the final form we arrange the
operators in the appropriate form, which is not always possible at
intermediate steps, but generally Dirac indices can be looked at, as
if they are incorporated into the coordinates.

Using the form of the $S$ matrix from Eq.~(\ref{Intro:SMatrix}) we
get
\begin{eqnarray}
\lefteqn{S(y_1,\cdots,y_N,y'_1,\cdots,y'_N) =}\qquad \nonumber\\
&&\left<0\right| :\bar\Psi(y'_1) \Psi(y_1)
  \cdots \bar \Psi(y'_N) \Psi(y_N) : \nonumber\\
&&\times{\cal T} \exp\left\{
  \int d^4x \  : \bar\Psi(x) [- i e \not\!\!A(x)] \Psi(x) :
  \right\} \left| 0 \right>
{}.
\end{eqnarray}
As the time, when we detect electrons and positrons, is $\infty$,
the time ordering can be extended to the whole expression.

We use now the Wick theorem in the following form
\cite{Bogoliubov:Shirkov:83}, where $O_i$ is an arbitrary fermionic
operator:
\begin{eqnarray}
\lefteqn{\left< 0 \right| {\cal T} O_1 \cdots O_M \left| 0 \right>
  =  \sum_{j, (j\not=i)} \mbox{sgn}(\pi(j,i))
  \left<0\right| {\cal T} O_j O_i \left|0\right>}\qquad\qquad \nonumber\\
&&\times  \left<0\right| {\cal T} O_1 \cdots \widehat{O_j} \cdots
\widehat{O_i} \cdots O_M \left|0\right>
{}.
\label{Wicktheorem}
\end{eqnarray}
Here $i$ is a fixed index, $\pi(j,i)$ is the permutation to put
$O_j$ and $O_i$ to the left of all other operators, and the hat in
the vacuum expectation value means that these operators have been
removed from the expression.  In our case the $O_i$ are either
$\bar\Psi$ or $\Psi$ and we can use the fact that
\begin{equation}
\left<0\right| {\cal T} \Psi(x) \Psi(y) \left|0\right> =
\left<0\right| {\cal T} \bar\Psi(x) \bar\Psi(y) \left|0\right> = 0
\end{equation}
and
\begin{equation}
\left<0\right| {\cal T} \Psi(x) \bar\Psi(y) \left|0\right>
  = i S_F(x-y)
,
\end{equation}
where $S_F$ is the usual Feynman propagator.  As we have also
normal-ordered products, the sum over $j$ in this case also excludes
those operators, which are in the same normal-ordered product as the
$O_i$.

In a first step we apply Eq.~(\ref{Wicktheorem}) to $\Psi(y_N)$ in
the expansion of $S_N$:
\widetext
\begin{eqnarray}
I_M &:=& \int dx_1 \cdots dx_M \left< 0 \right| {\cal T} :
  \bar\Psi(y_1') \Psi(y_1) \cdots \bar\Psi(y_N') \Psi(y_N):
  \nonumber\\
&&\times: \bar\Psi(x_1) \left[ - i e \not\!\!A(x_1) \right] \Psi(x_1):
  \cdots : \bar\Psi(x_M) \left[ - i e \not\!\!A(x_M) \right]\Psi(x_M):
  \left| 0 \right>\nonumber\\
&=& \int dx_1 \cdots dx_M \
   \sum_{i=1}^{M} \left< 0 \right| {\cal T} \Psi(y_N)
  \bar\Psi(x_i) \left| 0 \right>
  \left[ - i e \not\!\!A(x_i) \right]\nonumber\\
&&\times \left< 0 \right| {\cal T} : \bar\Psi(y_1') \Psi(y_1) \cdots
  \bar\Psi(y_N') : \Psi(x_i)
  :(1): \cdots \widehat{:(i):} \cdots :(M): \left| 0 \right>
  \nonumber\\
&=& M \int dz_1\ dx_1 \cdots dx_{M-1} \  S_F(y_N - z_1)
  e \not\!\!A(z_1)\nonumber\\
&&\times \left< 0 \right| {\cal T} : \bar\Psi(y_1') \Psi(y_1) \cdots
  \bar\Psi(y_N') : \Psi(z_1) :(1): \cdots :(M-1): \left| 0 \right>
,
\end{eqnarray}
where we have used the fact that each term in the sum is identical
after an even permutation and a renumbering of the variables $x_i$.
This integral is just one term in the expansion of the exponential
in $S$ and applying this reduction to every term we get
\begin{eqnarray}
S_N &=& \int dz_1 \  S_F(y_N - z_1) e \not\!\!A(z_1)
  \left< 0 \right| {\cal T} : \bar\Psi(y_1') \Psi(y_1) \cdots
  \bar\Psi(y_N') : \Psi(z_1) \nonumber\\
&&\times\exp\left\{ \int dx \  :\bar\Psi(x)
  \left[ - i e \not\!\!A(x) \right]
  \Psi(x) : \right\} \left| 0 \right>
{}.
\end{eqnarray}

In a next step we apply the Wick theorem to $\Psi(z_1)$ and get
similar as in the previous case for the integral
\begin{eqnarray}
I_M' &:=& \int dx_1 \cdots dx_M \left< 0 \right| {\cal T}
  :\bar\Psi(y_1') \Psi(y_1) \cdots \bar\Psi(y_N') : \Psi(z_1)
  :(1): \cdots :(M): \left| 0 \right> \nonumber\\
&=& \sum_{i=1}^{N} -i S_F(z_1-y_i') \int dx_1 \cdots dx_M
  \nonumber\\
&&\times \left< 0 \right| {\cal T}
  :\bar\Psi(y_1') \Psi(y_1) \cdots \widehat{\bar\Psi(y_i')}
  \Psi(y_i) \cdots \bar\Psi(y_N') :\
  :(1): \cdots :(M): \left| 0 \right> \nonumber\\
&&+ M \int dz_2\ dx_1\cdots dx_{M-1} \  S_F(z_1-z_2) e \not\!\!A(z_2)
  \nonumber\\
&&\times \left< 0 \right| {\cal T} : \bar\Psi(y_1') \Psi(y_1) \cdots
  \bar\Psi(y_N') : \Psi(z_2) :(1): \cdots :(M-1): \left| 0 \right>
,
\label{Pertubation:Wick2}
\end{eqnarray}
which is again a term in the expansion of $S_N$, so that our
amplitude is
\begin{eqnarray}
S_N &=& \sum_{i=1}^{N} -i \int dz_1 \  S_F(y_N - z_1) e \not\!\!A(z_1)
  S_F(z_1-y_i') \nonumber\\
&&\times \left< 0 \right| {\cal T}
  :\bar\Psi(y_1') \Psi(y_1) \cdots \widehat{\bar\Psi(y_i')}
  \Psi(y_i) \cdots \bar\Psi(y_N') : \nonumber\\
&&\times  \exp\left\{ \int dx \  :\bar\Psi(x)
  \left[ - i e \not\!\!A(x) \right]
  \Psi(x) : \right\} \left| 0 \right>\nonumber\\
&&+\int dz_1 dz_2 \  S_F(y_N - z_1) e \not\!\!A(z_1)
  S_F(z_1-z_2) e \not\!\!A(z_2) \nonumber\\
&&\times\left< 0 \right| {\cal T} : \bar\Psi(y_1') \Psi(y_1) \cdots
  \bar\Psi(y_N') : \Psi(z_2)
  \exp\left\{ \int dx \  :\bar\Psi(x)
  \left[ - i e \not\!\!A(x) \right]
  \Psi(x) : \right\} \left| 0 \right>
{}.
\end{eqnarray}

The vacuum expectation value in the first term is just the
$(N-1)$-pair creation amplitude, that is,
\begin{eqnarray}
S_N &=& \sum_{i=1}^{N} (-i) \int dz_1\  S_F(y_N - z_1) e \not\!\!A(z_1)
  S_F(z_1-y_i') \nonumber\\
&&\times \mbox{sgn}(\sigma)
  S_{N-1}(y_1,\cdots,y_{N-1},y_1',\cdots,\widehat{y_i'},\cdots,y_N')
  \nonumber\\
&&+\int dz_1 dz_2 \  S_F(y_N - z_1) e \not\!\!A(z_1)
  S_F(z_1-z_2) e \not\!\!A(z_2) \nonumber\\
&&\times\left< 0 \right| {\cal T} : \bar\Psi(y_1') \Psi(y_1) \cdots
  \bar\Psi(y_N') : \Psi(z_2)
  \exp\left\{ \int dx \  :\bar\Psi(x)
  \left[ - i e \not\!\!A(x) \right]
  \Psi(x) : \right\} \left| 0 \right>
,
\end{eqnarray}
where $\sigma$ is the permutation to rearrange the $\bar\Psi(y_j')$
and $\Psi(y_j)$ to the standard form.  We can use
Eq.~(\ref{Pertubation:Wick2}) again for the vacuum expectation value
in the second term.  Doing this recursively and assuming that the
series converges we finally get
\begin{eqnarray}
S_N &=& \sum_{i=1}^N \mbox{sgn}(\sigma)
  S_{N-1}(y_1,\cdots,y_{N-1},y_1',\cdots,\widehat{y_i'},\cdots,y_N')
  \nonumber\\
&&\times (-i) \biggl[ \int dz_1 \  S_F(y_N-z_1) e \not\!\!A(z_1)
  S_F(z_1-y_i') \nonumber\\
&&+ \int dz_1 dz_2\  S_F(y_N - z_1) e \not\!\!A(z_1) S_F(z_1 - z_2)
  e \not\!\!A(z_2) S_F(z_2-y_i') \nonumber\\
&&+ \cdots \biggr]
{}.
\end{eqnarray}

\narrowtext
The infinite series is just the normal perturbation theory expansion
for the pair creation without any disconnected parts.  We identify
it with the reduced one-pair creation amplitude $S^R(y_N,y_i')$.
Therefore we have found a way to reduce the $N$-pair amplitude to a
$(N-1)$-pair amplitude.  Recursively we can write for $S_N$:
\begin{eqnarray}
\lefteqn{S_N(y_1,\cdots,y_N,y_1',\cdots,y_N') =}\qquad \nonumber\\
&&  S_0 \sum_{\sigma} \mbox{sgn}(\sigma) S^R(y_1,y_{\sigma(1)}')
  \cdots S^R(y_N,y_{\sigma(N)}')
\label{Eq:reduction:pertubation}
\end{eqnarray}
with the vacuum amplitude
\begin{eqnarray}
S_0 &=& \left< 0 \right| {\cal T}
  \exp\left\{ \int dx  \ :\bar\Psi(x) [-i e \not\!\!A(x)] \Psi(x):
  \right\}\left| 0 \right> \nonumber\\
&=& \left< 0 \right| S \left| 0 \right>
{}.
\end{eqnarray}

This is identical to the result of the last section.  Therefore we
have found an expression for $S^R$, that is, for $s^{+-}$.  It is
just the perturbation theory expansion of a fermion interacting an
arbitrary number of times with an external field without any
disconnected parts.

Finally we want to find also a perturbation theory expression for
$S_0$.  Expanding the exponential we get
\begin{equation}
S_0 = \sum_{M=0}^{\infty} \frac{1}{M!} \tilde I_M
,
\end{equation}
with $\tilde I_M$ defined as
\begin{eqnarray}
\lefteqn{\tilde I_M := \int dx_1 \cdots dx_M}\ \  \nonumber\\
&&\times \left< 0 \right | {\cal T}
  : \bar\Psi(x_1) \left[ - i e \not\!\!A(x_1) \right] \Psi(x_1) :
  \cdots : (M): \left| 0 \right >
{}.
\end{eqnarray}

We are now using the Wick theorem again in order to reduce this
expression.  Applying it to $\Psi(x_1)$ we get
\begin{eqnarray}
\tilde I_M &=& (-i) (M-1) \int dx_1 dx_2 \cdots dx_M \nonumber\\
&&\times  e \not\!\!A(x_1) S_F(x_1 - x_2) e \not\!\!A(x_2)
  \nonumber\\
&&\times  \left< 0 \right | {\cal T}  \bar\Psi(x_1) \Psi(x_2)
  :(3): \cdots :(M): \left| 0 \right >
{}.
\end{eqnarray}
Similar we get for the following expression
\widetext
\begin{eqnarray}
\tilde J_M(x_1,x_2) &:=& \int dx_3 \cdots dx_M
  \left< 0 \right | {\cal T} \bar\Psi(x_1) \Psi(x_2)
  : \bar\Psi(x_3) \left[ - i e \not\!\!A(x_3) \right] \Psi(x_3) :
  \cdots :(M): \left| 0 \right >
  \nonumber\\
&=& -i S_F(x_2 - x_1)   \int dx_3 \cdots dx_M
  \left< 0 \right | {\cal T}  :(3) : \cdots
  :(M):  \left| 0 \right > \nonumber\\
&&+ (M-2) \int dx_3 dx_4\cdots dx_M\
   i S_F(x_2 - x_3) \left[- i e \not\!\!A(x_3)\right] \nonumber\\
&&\times  \left< 0 \right | {\cal T}  \bar\Psi(x_1) \Psi(x_3)
  :(4): \cdots :(M): \left| 0 \right >
{}.
\end{eqnarray}
That is,
\begin{equation}
\tilde I_M = -i (M-1)\int dx_1 dx_2\  e \not\!\!A(x_1)
  S_F(x_1 - x_2) e \not\!\!A(x_2) \tilde J_M(x_1,x_2)
\end{equation}
and
\begin{equation}
\tilde J_M(x_1,x_2) = -i S_F(x_2 - x_1) \tilde I_{M-2} +
  (M-2) \int dx_3 \  S_F(x_2- x_3) e \not\!\!A(x_3)
  \tilde J_{M-1}(x_1,x_3)
{}.
\end{equation}

\narrowtext
Using both relations in order to reduce $\tilde I_M$ in terms of
$\tilde I_k$ we get
\begin{eqnarray}
\tilde I_M &=& (M-1)\  C_2 \tilde I_{M-2} + (M-1)(M-2)\
  C_3 \tilde I_{M-3} \nonumber\\
&&+ \cdots + (M-1)\cdots 2\  C_{M-1} \tilde I_1 \nonumber\\
&&  + (M-1) \cdots 2\  1\  C_M \tilde I_0
,
\end{eqnarray}
with $C_N$ defined as
\begin{eqnarray}
C_N &:=& (-1) \int dx_1 \cdots  dx_N \nonumber\\
&&  \mbox{Tr} \biggl[ e \not\!\!A(x_1)
  S_F(x_1 - x_2) e \not\!\!A(x_2) \cdots \nonumber\\
&&\times e \not\!\!A(x_N)
  S_F(x_N - x_1) \biggr]
,
\end{eqnarray}
writing explicitly the Dirac indices to get the trace.  For ease of
writing we define also $C_0 = C_1 = 0$ and use that $\tilde I_1 = 0$
and $\tilde I_0 = 1$.  With these we can write more compactly
\begin{eqnarray}
\tilde I_M &=& \sum_{k=1}^{M} C_k \tilde I_{M-k}
  \frac{(M-1)!}{(M-k)!} \nonumber\\
&=& \frac{M!}{M} \sum_{k=1}^{M} C_k \frac{\tilde I_{M-k}}{(M-k)!}
{}.
\label{Vac:reduction}
\end{eqnarray}
This reduction formula can now be used in order to find an
expression for $S_0$.  For this we write $\tilde I_M$ in a series
ordered by the number of $C_i$'s (that is, as we will see later, the
number of loops) that appear in each term after a complete
reduction.  We see that the formula above just gives us a recipe,
how the next order can be derived from the lower one.  Only one term
with no $C_i$ in it exists:
\begin{equation}
\tilde I_0^{(0)} = 1
,
\end{equation}
giving therefore in zeroth order for $S_0$:
\begin{equation}
S_0^{(0)} = 1
{}.
\end{equation}
Using this only nonzero $\tilde I_M^{(0)}$ in
Eq.~(\ref{Vac:reduction}) we get in first order only a contribution
if $k=M$
\begin{equation}
\tilde I_M^{(1)} = \frac{M!}{M} C_M
\end{equation}
and for $S_0$
\begin{equation}
S_0^{(1)} = \sum_{M=0}^{\infty} \frac{1}{M!} \tilde I_M^{(1)}
  = \sum_{M=1}^{\infty} \frac{C_M}{M}
{}.
\end{equation}

In second order we get
\begin{eqnarray}
\tilde I_M^{(2)} &=& \frac{M!}{M} \sum_{k=1}^{M} C_k
  \frac{\tilde I_{M-k}^{(1)}}{(M-k)!}
  = \frac{M!}{M} \sum_{k=1}^{M} k \frac{C_k}{k}
  \frac{C_{M-k}}{M-k} \nonumber\\
&&= \frac{M!}{M} \frac{1}{2} \sum_{k=1}^{M}
  \left[ k \frac{C_k}{k} \frac{C_{M-k}}{M-k} +
  (M-k) \frac{C_{M-k}}{M-k} \frac{C_k}{k} \right] \nonumber\\
&&= \frac{M!}{2} \sum_{k=1}^{M} \frac{C_k}{k}
  \frac{C_{M-k}}{M-k}
,
\end{eqnarray}
where we have changed the summation index from $k$ to $M-k$ in one
step.  For $S_0$ we get
\begin{eqnarray}
S_0^{(2)} &=& \sum_{M=0}^{\infty} \frac{1}{M!} \tilde I_M^{(2)}
  = \frac{1}{2} \sum_{M=0}^{\infty} \sum_{k=1}^{M} \frac{C_k}{k}
  \frac{C_{M-k}}{M-k} \nonumber\\
  &=& \frac{1}{2} \left[\sum_{k=1}^{\infty} \frac{C_k}{k} \right]
  \left[\sum_{l=1}^{\infty} \frac{C_l}{l} \right]
  = \frac{1}{2} \left[S_0^{(1)} \right]^2
{}.
\end{eqnarray}
Using the same transformations we can express also all higher terms
of $S_0$, for which we get
\begin{equation}
S_0^{(n)} = \frac{1}{n!} \left[S_0^{(1)} \right]^n
,
\end{equation}
so that we finally get the vacuum amplitude as
\begin{eqnarray}
\lefteqn{S_0 = \exp\left[S_0^{(1)} \right] =
  \exp\biggl\{- \sum_{M=2}^{\infty} \frac{1}{M} \int dx_1 \cdots
  dx_M}\quad \nonumber\\
  &&\times \mbox{Tr}
  \left[ e \not\!\!A(x_1) S_F(x_1-x_2) \cdots
  e \not\!\!A(x_M) S_F(x_M-x_1)\right] \biggr\}
{}.
\nonumber\\
\end{eqnarray}
This is a well known result
\cite{Schwinger:54,Itzykson:Zuber:80,Bialynicki-Birula:75,Feynman:QED}.
 Its interpretation is straightforward.  The sum in the exponential
is just the sum over all single loops, whereas the exponential
accounts for the fact that we can have also two or more loops.

With this we have found now perturbation-theory expressions for all
terms occurring in the reduction formula
Eq.~(\ref{Eq:reduction:pertubation}), that is, for the reduced
one-pair amplitude and the vacuum amplitude.

A final remark has to be made about the vacuum amplitude $S_0$.
Already Schwinger found out that the expression for $S_0$ is not
finite due to infinities in the imaginary part of $S_0^{(1)}$
\cite{Schwinger:54}(see also \cite{Schweber:61}).  But he also
mentioned that in the total probabilities we need only the absolute
value squared of $S_0$, where only the real part of $S_0^{(1)}$
contributes, which is finite.

\section{Comparison with earlier results}
\label{Sec:Comparision}

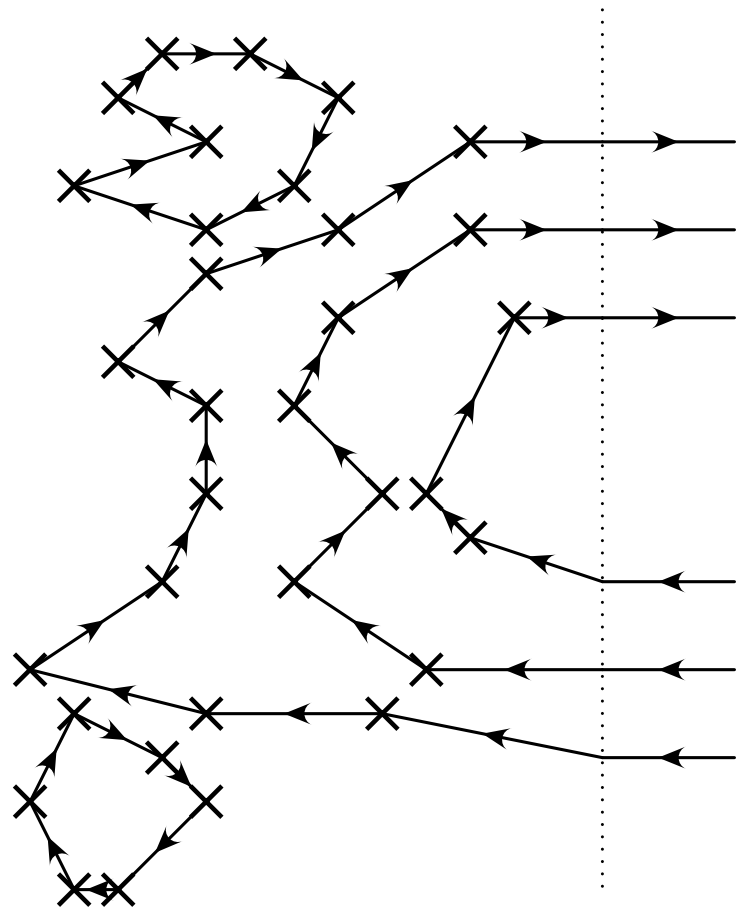
\begin{figure}
\caption{Graphical illustration of the general form of
the $N$-pair production process. The creation of a pair is
described by a connected fermion line entering from and leaving to
the future. The vacuum processes are described by all sorts of
closed fermion loops. The interaction with the external field is
shown as a cross.}
\label{Fig:GenN}
\end{figure}

The results of the previous two sections suggest the following
picture of the pair production: In the Feynman picture the $N$-pair
production can be described by two forms of processes
(Fig.~\ref{Fig:GenN}).  $N$ positron lines enter the interaction
region coming from the future.  They interact with the external
field an arbitrary number of times, where they may change also their
direction in time.  Finally they leave the interaction region as
electron lines moving into the normal time direction.  Besides these
processes, which are characterized each by a continuous line coming
{}from and leaving to the future, there are also processes, which
consists of closed loops.  As they remain entirely inside the
interaction region, they are not visible as physical processes.
These closed loops form the vacuum amplitude.  From this picture it
is also clear that the vacuum amplitude is the same for all $N$-pair
amplitudes and therefore a common factor in all of them.  The fact
that we are dealing with multiple particles only shows up in the
calculation through the antisymmetrisation with respect to all
electrons (or equivalent to all positrons).

This means that for a complete calculation of the $N$-pair creation
probabilities we need to know only the reduced one-pair creation
amplitude together with the vacuum amplitude.  Both can be
calculated in principle using perturbation theory, where the reduced
one-pair amplitude is identical to the usual perturbation series
result and setting the vacuum amplitude to one, that is, neglecting
all diagrams with disconnected parts.

If one neglects the exchange terms in the calculation of the
probability, it suffices to know the reduced one-pair creation
probability, as this is the only result needed for the Poisson
distribution of the multiple-pair probabilities.

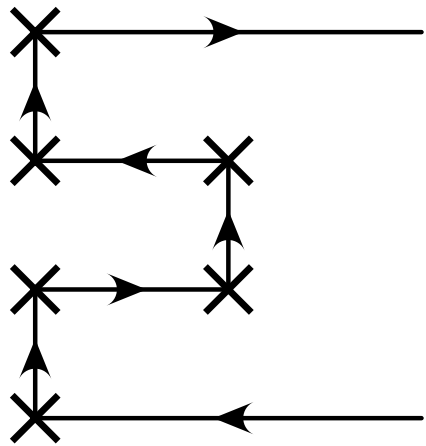
\begin{figure}
\caption{One possible higher-order pair creation process, where
the Pauli principle is neglected in the intermediate state.
Interaction with the external field is shown as a cross.}
\label{Fig:PViol}
\end{figure}

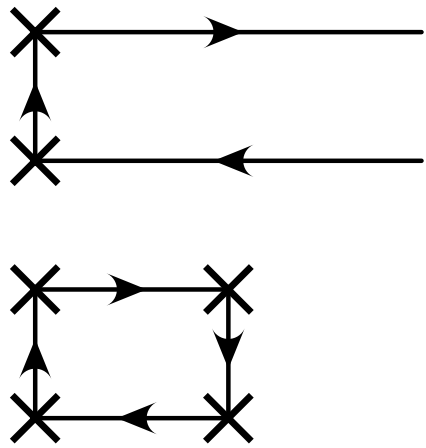
\begin{figure}
\caption{Vacuum correction to the one-pair creation connected
by an exchange of two lines with the process shown in
Fig.~\protect\ref{Fig:PViol}.}
\label{Fig:LVac}
\end{figure}

Note that according to the rules of the Feynman diagrams we have
neglected the antisymmetrisation and therefore the Pauli principle
for all intermediate states.  There is a deeper reason for doing so,
which has to do with the connection of the higher-order processes
and the occurrence of the vacuum processes through
antisymmetrisation.  This is, for example, discussed in detail in
\cite{Bialynicki-Birula:75,Feynman:QED}.  Let us look, for example,
at a typical higher-order diagram in perturbation theory, where we
would expect corrections because of the Pauli principle.  In
Fig.~\ref{Fig:PViol} we would expect a deviation as we have two
electrons, which are not allowed to be in the same state.  (The same
is also true for the two positrons.) But it can be shown that the
contributions from this process are just canceled by those of the
vacuum correction shown in Fig.~\ref{Fig:LVac}, where the electron
in the loop and the produced electron are not allowed to be in the
same state also.  Both processes are connected with each other, as
we get one from the other, if we exchange the two electron lines or
the two positron lines.  This result is of a very general nature, so
that the antisymmetrisation of all intermediate particles can be
dropped, if we also include the vacuum processes (again without
antisymmetrisation).

This shows that the higher-order multiple-particle corrections to
the one-pair creation are connected with the vacuum process.
Therefore we think that these processes should be studied in more
detail.  Especially as we know that the vacuum processes are not
negligible, as they decrease the reduced probabilities larger than
one to values less than one.

The inclusion of these higher-order processes is straightforward in
our model.  As the Poisson distribution needs only the neglection of
the exchange terms, we can include the higher-order terms into the
calculation of the reduced probability and use this result then in
the Poisson distribution for the multiple-pair probabilities.  This
procedure is consistent, as it includes automatically the
higher-order corrections to the vacuum amplitude.  Therefore the
calculation changes only the reduced probability used in the Poisson
distribution but not the explicit form.

Let us compare this result with the other models
\cite{Baur:Unitarity:90,Rhoades-Brown:91,Best:92}: All three models
get a Poisson distribution for the multiple-pair creation as well
but only based on the summation of a restricted class of diagrams.
Studying further these approximations one finds that all of them are
essentially ``quasi boson'' approximations.  This means that they
have as fundamental processes a pair creation and a pair
annihilation process (neglecting at the moment the Coulomb
scattering).  Combining these processes all previous models assume
that only pairs, which have been produced together in a creation
process, can be annihilated.  Therefore the electron and the
positron are seen as an unbreakable pair, which behaves more or less
like a boson.  And for bosons in an external field the Poisson
distribution comes out exactly \cite{Itzykson:Zuber:80}.  Even the
inclusion of Coulomb scattering in some of these models does not
change the ``quasi bosonic'' nature of them.  Our calculation
indicates that there are higher-order processes in which electrons
and positrons do not behave as ``quasi bosons''.  These processes
are multiple-particle processes in the sense that more than one
electron or positron are essentially needed in an intermediate step.

The fact that these are multiple-particle effects also shows the
advantage of using the Feynman boundary conditions instead of the
retarded boundary conditions used in the Dirac sea picture.  In the
Feynman picture the production of a pair is described by a
continuous fermion line, where the inclusion of multiple-particle
effects presents no difficulties.  The advantage of the Dirac sea
picture is mainly that its particle hole interpretation allows to
treat the creation of one electron-positron pair in a single
particle formalism.  As the higher-order processes need the
existence of more than one electron and one hole, this advantage of
the Dirac sea picture is no longer existent then.

The other advantage of the Feynman boundary conditions is that the
different processes can be separated from each other.  Especially
the vacuum processes are independent of the reduced pair creation
processes.  This is not the case in the Dirac sea picture, where
vacuum processes and pair creation processes can not be separated in
the intermediate steps but only in the final state.

In the following we show that the processes of the type of
Fig.~\ref{Fig:PViol} are the lowest-order corrections to the
one-pair creation processes if we neglect Coulomb scattering terms.
For this we restrict ourself to second-order Magnus theory.

\section{The $S$ operator in Magnus theory}
\label{Sec:Magnus}

The Magnus theory can be seen as an expansion in the interaction
time.  For large values of $\gamma$ the interaction time of the two
heavy-ion fields is very short, therefore the use of this
approximation seems to be justified.  Up to second order the $S$
operator is given by \cite{Bialynicki-Birula:69,Pechukas:Light:66}
\widetext
\begin{eqnarray}
S &=& {\cal T} \exp\left[-i \int_{-\infty}^{+\infty} H_I(t)
  \ dt \right] \nonumber\\
&\approx& \exp\left[-i \int_{-\infty}^{+\infty} H_I(t) dt +
  \frac{1}{2} (-i)^2 \int_{-\infty}^{+\infty} dt_2
  \int_{-\infty}^{t_2} dt_1 \left[ H_I(t_2), H_I(t_1)\right] +
  \cdots\right]
{}.
\end{eqnarray}

The commutator of the $H_I$'s for $t_1<t_2$ is the difference of the
time-ordered and the anti-time-ordered product.  We rewrite $S$ as
\begin{eqnarray}
  S &=& \exp\biggl\{-i \int_{-\infty}^{+\infty} H_I(t) dt +
    \frac{1}{4} (-i)^2 \int_{-\infty}^{+\infty} dt_2
    \int_{-\infty}^{+\infty} dt_1 \nonumber\\
  &&\times {\cal T}\left[ H_I(t_2) H_I(t_1) \right]
    - {\cal A} \left[ H_I(t_2) H_I(t_1) \right] +
    \cdots\biggr\}\\
&=& \exp\biggl\{\int d^4x \
  e : \bar\Psi(x) [- i e \not\!\!A(x)] \Psi(x) : \nonumber\\
&&+ \frac{1}{4} \int d^4x_1\ d^4x_2\ ({\cal T}-{\cal A})
  : \bar\Psi(x_2) [- i e \not\!\!A(x_2)] \Psi(x_2) :\
  : \bar\Psi(x_1) [- i e \not\!\!A(x_1)] \Psi(x_1) : \biggr\}
{}.
\end{eqnarray}

\narrowtext
Now we use again the Wick theorem in order to put the field
operators into normal-ordered form.  We use it for the time-ordered
and anti-time-ordered products in the form
\begin{mathletters}
\begin{eqnarray}
{\cal T} A B &=& : A B : + \left<0| {\cal T} A B | 0 \right>
  =\  : A B : + \left<{\cal T} A B \right>,\\
{\cal A} A B &=& : A B : + \left<0| {\cal A} A B | 0 \right>
  =\  : A B : + \left<{\cal A} A B \right>
{}.
\end{eqnarray}
\end{mathletters}
As the first term in $S$ is in normal-ordered form already, only the
second one has to be rearranged.  For the time-ordered product we
get
\widetext
\begin{eqnarray}
{\cal T} : \bar\Psi_2 \Psi_2 :\ : \bar\Psi_1 \Psi_1 : &=&
  : \bar\Psi_2 \Psi_2 \bar\Psi_1 \Psi_1 : +
  : \Psi_2 \bar\Psi_1 : \left<{\cal T} \bar\Psi_2 \Psi_1 \right>
  \nonumber\\
&&+ : \bar\Psi_2 \Psi_1 : \left<{\cal T} \Psi_2 \bar\Psi_1 \right> +
  \left< {\cal T}\bar\Psi_2 \Psi_1 \right>
  \left<{\cal T}\Psi_2 \bar\Psi_1 \right>
,
\end{eqnarray}
and the same for the anti-time-ordered product by replacing
${\cal T}$ with ${\cal A}$. We get for $S$
\begin{eqnarray}
S &=& \exp\biggl\{ \int dx\ [-ie \not\!\!A(x)_{\alpha\beta}(x)]
  : \bar\Psi_\alpha \Psi_\beta : +
  \frac{1}{4} \int dx_1 dx_2\ [-ie \not\!\!A(x_2)_{\alpha\beta}]
  [-ie \not\!\!A(x_1)_{\gamma\delta}] \nonumber\\
&&\times \bigl[: \Psi_\beta(x_2) \bar\Psi_\gamma(x_1) :
  \left<{\cal D} \bar\Psi_\alpha(x_2) \Psi_\delta(x_1) \right> +
  : \bar\Psi_\alpha(x_2) \Psi_\delta(x_1) :
  \left<{\cal D} \Psi_\beta(x_2) \bar\Psi_\gamma(x_1) \right>
  \nonumber\\
&&+ \left< {\cal T}\bar\Psi_\alpha(x_2) \Psi_\delta(x_1) \right>
  \left<{\cal T}\Psi_\beta(x_2) \bar\Psi_\gamma(x_1) \right>
  - \left< {\cal A}\bar\Psi_\alpha(x_2) \Psi_\delta(x_1) \right>
  \left<{\cal A}\Psi_\beta(x_2) \bar\Psi_\gamma(x_1) \right>
  \bigr]\biggr\}
,\nonumber\\
\label{Eq:SMatrixTA}
\end{eqnarray}
where we have defined $\cal D$ as ${\cal T} - {\cal A}$.
 The last two terms can be rewritten to give
\begin{equation}
\left< {\cal D}\bar\Psi_\alpha(x_2) \Psi_\delta(x_1) \right>
  \left<{\cal T}\Psi_\beta(x_2) \bar\Psi_\gamma(x_1) \right>
  + \left< {\cal A}\bar\Psi_\alpha(x_2) \Psi_\delta(x_1) \right>
  \left<{\cal D}\Psi_\beta(x_2) \bar\Psi_\gamma(x_1) \right>
{}.
\end{equation}
Exchanging the integration of $dx_1$ and $dx_2$ we see that the
second and the third term of Eq.~(\ref{Eq:SMatrixTA}) can be
combined and also the last two terms, so that we get
\begin{eqnarray}
S &=& \exp\biggl\{ \int dx\ [-ie \not\!\!A(x)_{\alpha\beta}(x)]
  : \bar\Psi_\alpha \Psi_\beta : +
  \frac{1}{4} \int dx_1 dx_2\ [-ie \not\!\!A(x_2)_{\alpha\beta}]
  [-ie \not\!\!A(x_1)_{\gamma\delta}] \nonumber\\
&&\times \biggl[2 : \Psi_\beta(x_2) \bar\Psi_\gamma(x_1) :
  \left<{\cal D} \bar\Psi_\alpha(x_2) \Psi_\delta(x_1) \right>
  \nonumber\\
&&+ \left<({\cal T}+{\cal A})\Psi_\beta(x_2)
  \bar\Psi_\gamma(x_1) \right>
  \left< {\cal D}\bar\Psi_\alpha(x_2) \Psi_\delta(x_1) \right>
  \biggr]\biggr\}
{}.
\label{Eq:SMatrixx}
\end{eqnarray}

\narrowtext
Now we transform $S$ into momentum space.  For this we need the
Fourier transform of the field operator as well as of the vacuum
expectation values.

The field operator and its conjugate can be written as
\begin{mathletters}
\begin{eqnarray}
\Psi(x) &=& \sum_s \int d\tilde p
  \biggl[ b(p,s) u(p,s) \exp(-ipx) \nonumber\\
&&  + d^+(p,s) v(p,s) \exp(ipx)\biggr],\\
\bar\Psi(x) &=& \sum_s \int d\tilde p
  \biggl[ b^+(p,s) \bar u(p,s) \exp(ipx) \nonumber\\
&&  + d(p,s) \bar v(p,s) \exp(-ipx)\biggr]
{}.
\end{eqnarray}
\label{Magnus:field}
\end{mathletters}
where we have introduced $d\tilde p$, which is defined as
\begin{equation}
d\tilde p := \frac{d^3 p}{(2\pi)^{3/2} (2 p_0)^{1/2}}
{}.
\end{equation}
In the following we need also the factor of $d\tilde p$ alone
together with the Lorentz invariant phase space, therefore we define
also
\begin{equation}
\gamma(p) := \frac{1}{(2\pi)^{3/2} (2 p_0)^{1/2}}
,
\end{equation}
and
\begin{equation}
d\Gamma(p) := \frac{d^3 p}{(2\pi)^3 2 p_0}
{}.
\end{equation}

With the usual relations between vacuum expectation values,
propagators, and singular function, one gets for the vacuum
expectation values
\cite{Bjorken:Drell:66,Bjorken:Drell:67,Jauch:Rohrlich:76}
\begin{mathletters}
\begin{eqnarray}
\lefteqn{\left< 0 \right| {\cal T} \Psi(x) \bar \Psi(x')
  \left| 0 \right> = i S_F(x-x')}\qquad \nonumber\\
&&= i \int \frac{d^4p}{(2\pi)^4}
    \frac{\not\!p + m}{p^2 -m^2 + i \epsilon} \exp(- i p (x-x'))
,
\end{eqnarray}
\begin{eqnarray}
\lefteqn{\left< 0 \right| {\cal A} \Psi(x) \bar \Psi(x')
  \left| 0 \right> = - i S_A(x-x') }\qquad\nonumber\\
&&= - i \int \frac{d^4p}{(2\pi)^4}
    \frac{\not\!p + m}{p^2 -m^2 - i \epsilon} \exp(- i p (x-x'))
{}.
\end{eqnarray}
\end{mathletters}

Combining both we get for ${\cal D}$ and ${\cal T}+{\cal A}$:
\begin{eqnarray}
\lefteqn{\left< 0 \right| {\cal D} \Psi(x) \bar \Psi(x')
  \left| 0 \right> = i \left[ S_F(x-x') + S_A(x-x') \right] }
  \quad \nonumber\\
&&= i \int \frac{d^4p}{(2\pi)^4} (\not\!p + m)\left(
  \frac{1}{p^2 - m^2 + i \epsilon}
  + \frac{1}{p^2 -m^2 - i \epsilon}\right) \nonumber\\
&&\times \exp(- i p (x-x')) \nonumber\\
&&= 2 i \int \frac{d^4p}{(2\pi)^4} (\not\!p + m)
  \frac{\mbox{P.P.}}{p^2 -m^2} \exp(- i p (x-x'))
,
\end{eqnarray}
where P.P.  denotes the principal part of the integral, and
\begin{eqnarray}
\lefteqn{\left< 0 \right| ({\cal T}+{\cal A}) \Psi(x) \bar\Psi(x')
\left| 0 \right> =  i \left[S_F(x-x') - S_A(x-x')\right] }
  \qquad \qquad \nonumber\\
&=& \int d\Gamma(p)\  (\not\!p + m)
 \exp(- i p ( x - x')) \nonumber\\
&&  - \int d\Gamma(p)\  (\not\!p - m) \exp(i p (x-x'))
,
\end{eqnarray}
which describes the propagation of on-shell electrons and positrons.

These forms of the vacuum expectation values are put into $S$
together with the decomposition of the field operators $\Psi(x)$ and
$\bar\Psi(x)$ and using also the Fourier transform of the external
field
\begin{equation}
A(x) = \int \frac{d^4q}{(2\pi)^4} A(q) \exp(- i q x)
{}.
\end{equation}

The integration over the coordinate space can be done, giving
deltafunctions for the momenta.  Reducing the momentum integrals
with the help of these, we finally get
\widetext
\begin{eqnarray}
S &=&
  \exp \Biggl\{ -i e \int d\tilde p_1\ d\tilde p_2 \nonumber\\
&&\times\biggl[ : b^+(p_1) b(p_2) :\ \bar u(p_1)
  \not\!\!A(p_1-p_2) u(p_2) \nonumber\\
&&+ : b^+(p_1) d^+(p_2) :\ \bar u(p_1) \not\!\!A(p_1+p_2) v(p_2)
  \nonumber\\
&&+ : d(p_1) b(p_2) :\ \bar v(p_1) \not\!\!A(-p_1-p_2) u(p_2)
  \nonumber\\
&&+ : d(p_1) d^+(p_2) :\ \bar v(p_1) \not\!\!A(-p_1+p_2) v(p_2)
 \biggr] \nonumber\\
&& -i e^2 \int d\tilde p_1\ d\tilde p_2 \ \frac{dp}{(2\pi)^4}
  \nonumber\\
&&\times \biggl[
  : b^+(p_1) b(p_2) :\ \bar u(p_1) \not\!\!A(p_1-p)
  \left(\not\!p + m \right)
  \frac{\mbox{P.P.}}{p^2-m^2} \not\!\!A(p-p_2) u(p_2) \nonumber\\
&&+ : b^+(p_1) d^+(p_2):\ \bar u(p_1) \not\!\!A(p_1-p)
  \left(\not\!p + m \right)
  \frac{\mbox{P.P.}}{p^2-m^2} \not\!\!A(p+p_2) v(p_2) \nonumber\\
&&+ : d(p_1) b(p_2):\ \bar v(p_1) \not\!\!A(-p_1-p)
  \left(\not\!p + m \right)
  \frac{\mbox{P.P.}}{p^2-m^2} \not\!\!A(p-p_2) u(p_2) \nonumber\\
&&+ : d(p_1) d^+(p_2):\ \bar v(p_1) \not\!\!A(-p_1-p)
  \left(\not\!p + m \right)
  \frac{\mbox{P.P.}}{p^2-m^2} \not\!\!A(p+p_2) v(p_2) \biggr]
  \nonumber\\
&& + i \frac{e^2}{2} \int \frac{dp}{(2\pi)^4}
    d\Gamma(p')\
  \mbox{Tr}\left[ (\not\!p'+m) \not\!\!A(p'-p)
  (\not\!p+m) \frac{\mbox{P.P.}}{p^2-m^2} \not\!\!A(p-p') \right]
  \nonumber\\
&& - i \frac{e^2}{2} \int \frac{dp}{(2\pi)^4}
   d\Gamma(p')\
  \mbox{Tr}\left[ (\not\!p'-m) \not\!\!A(-p'-p)
  (\not\!p+m) \frac{\mbox{P.P.}}{p^2-m^2} \not\!\!A(p+p') \right]
  \Biggr\}
{}.
\label{Eq:SopMagnus}
\end{eqnarray}

\narrowtext
Again, we have not written explicitly the spins of the leptons,
which we look at as included in the momenta.

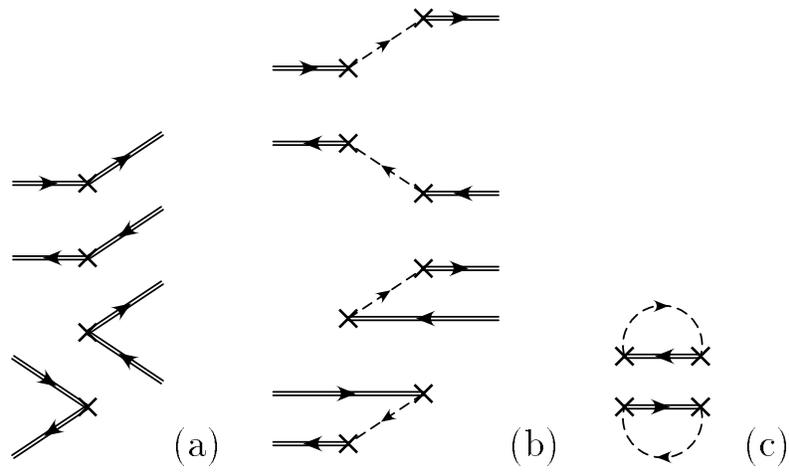
\begin{figure}
\caption{Graphical illustration of the processes occurring in first
(a) and second-order Magnus theory (b and c). Dotted lines
denote the propagator, where only the principal part is taken,
double lines denote electrons or positrons, which are on-shell.
Interaction with the external field is shown as a cross.}
\label{Fig:Magnus}
\end{figure}

The interpretation of the individual terms is straightforward (see
Fig.~\ref{Fig:Magnus}).  Terms of the form $:b^+b:$ are electron
Coulomb scattering terms, $:d d^+ :$ the corresponding positron
scattering terms.  $:b^+ d^+:$ corresponds to pair creation and $:d
b:$ to pair annihilation.  The last two terms have no operators in
them.  They correspond to the lowest-order vacuum corrections.  As
they are only numbers, they commute with all other terms.

\section{Application to electromagnetic pair creation}
\label{Sec:Highercorr}

We apply this result, which is true for an arbitrary external field,
to the special case of pair production in heavy-ion collisions.  In
this case there are some kinematical restrictions, so that some of
the terms in $S$ can be dropped.  Especially there is no pair
creation or pair annihilation in first-order Magnus theory and we
can replace the principal value integrals in the pair creation and
annihilation term in the second order by ordinary integrals, as the
intermediate state is not allowed to be on-shell in this case.

As we want to concentrate on the multiple-particle effects we will
neglect also all Coulomb scattering terms.  It is generally assumed
that these terms do change the differential probabilities, but are
only of minor importance for the total probabilities.  Also their
influence seems to become smaller for higher energies.  Therefore we
will neglect all terms of the form $: b^+ b:$ and $:d d^+:$.  Also
the vacuum terms will be dropped, as we are only interested in the
calculation of the reduced amplitude, where vacuum corrections do
not appear.  The $S$ operator with these approximations is
\begin{eqnarray}
S &=& \exp\biggl\{ -i e^2 \int d\tilde p_1\ d\tilde p_2
  \frac{dp}{(2\pi)^4} \nonumber\\
&&\times \biggl[  b^+(p_1) d^+(p_2)\nonumber\\
&&\times \bar u(p_1) \not\!\!A(p_1-p)
  \frac{\not\!p + m}{p^2-m^2} \not\!\!A(p+p_2) v(p_2) \nonumber\\
&&+  d(p_1) b(p_2)\nonumber\\
&&\times \bar v(p_1) \not\!\!A(-p_1-p)
  \frac{\not\!p + m}{p^2-m^2} \not\!\!A(p-p_2) u(p_2)\biggr]\biggr\}
{}.
\end{eqnarray}

The integrals over $p$ in this expression are just the pair creation
and pair annihilation amplitude in second-order Born approximation.
We define the pair creation and annihilation potential $V$ and $V^*$
as
\begin{mathletters}
\begin{eqnarray}
\lefteqn{V(p_1,p_2) =
e^2 \int \frac{dp}{(2\pi)^4}}\qquad \nonumber\\
&&\times \bar u(p_1) \not\!\!A(p_1-p)
  \frac{\not\!p + m}{p^2-m^2} \not\!\!A(p+p_2) v(p_2) \\
\lefteqn{V^*(p_2,p_1) =
e^2 \int \frac{dp}{(2\pi)^4}}\qquad  \nonumber\\
&&\times \bar v(p_1) \not\!\!A(-p_1-p)
  \frac{\not\!p + m}{p^2-m^2} \not\!\!A(p-p_2) u(p_2)
\end{eqnarray}
\end{mathletters}
(where $V^*$ is just the complex conjugate of $V$)
in order to write $S$ as:
\begin{eqnarray}
S &=& \exp\Biggl\{-i \int d\tilde p\ d\tilde q \nonumber\\
&&\times  \left[ b^+(p) d^+(q) V(p,q) + d(q) b(p) V^*(p,q) \right]
  \Biggr\}
{}.
\end{eqnarray}
The matrix element of the pair creation in second order is connected
with the potential $V$ through
\begin{equation}
M(p,q) = -i V(p,q)
{}.
\end{equation}

For the reduced one-pair creation amplitude we have to calculate
\begin{equation}
\left< f \right| S \left| i \right>
  = \left< 0 \right| d(q_f) b(p_f) S \left| 0 \right>
{}.
\end{equation}

We are now expanding the exponential in $S$ in order to get the
different contributions to the one-pair creation.  In lowest order
we expect to get back the second-order Born result.  We get
\widetext
\begin{eqnarray}
\left< f \right| S^{(1)} \left| i \right> &=&
  -i \int d\tilde p\ d\tilde q\
  V(p,q) \left< 0 \right| d(q_f) b(p_f) b^+(p) d^+(q) \left| 0
  \right>\nonumber\\
&=& - i \gamma(p_f) \gamma(q_f) V(p_f,q_f)\nonumber\\
&=& M(p_f,q_f) \gamma(p_f) \gamma(q_f)
,
\end{eqnarray}

The differential probability in first order therefore is
\begin{equation}
P(p_f,q_f) = \left| M(p_f,q_f) \right|^2 \gamma^2(p_f) \gamma^2(q_f)
\end{equation}
and the total probability is given by
\begin{equation}
P_{\rm total} = \int \left| M(p_f,q_f) \right|^2 d\Gamma(p_f)
  d\Gamma(q_f)
,
\end{equation}
which is indeed the Born result.

As the vacuum expectation values, we get from higher orders, vanish,
if the number of creation and annihilation operators is not equal
for each kind of particles, we see easily that only odd orders in
the expansion of $S$ contribute.  The next order is therefore the
third one, where we get
\begin{eqnarray}
\lefteqn{\left< f \right| S^{(3)} \left| i \right> = \frac{i}{3!} \int
d\tilde p_1\ d\tilde p_2\ d\tilde p_3\ d\tilde q_1\ d\tilde q_2\
  d\tilde q_3 }\qquad \nonumber\\
&&\times  \biggl[ V(p_1,q_1) V^*(p_2,q_2) V(p_3,q_3) \left< 0 \right|
d(q_f) b(p_f) b^+(p_1) d^+(q_1) d(q_2) b(p_2) b^+(p_3) d^+(q_3)
\left| 0 \right> \nonumber\\
&& + V^*(p_1,q_1) V(p_2,q_2) V(p_3,q_3)  \left< 0 \right|
d(q_f) b(p_f) d(q_1) b(p_1) b^+(p_2) d^+(q_2) b^+(p_3) d^+(q_3)
\left| 0 \right> \biggr]
.\nonumber\\
\end{eqnarray}
Here we have used the fact that only two of the eight possible
combinations do not vanish trivially.  Again the vacuum expectation
value can be calculated using the anticommutation relations of the
particle operators.  The first expectation value is
\begin{eqnarray}
\lefteqn{\left< 0 \right| d(q_f) b(p_f) b^+(p_1) d^+(q_1) d(q_2)
  b(p_2) b^+(p_3) d^+(q_3) \left| 0 \right> =}\qquad\nonumber\\
&& \delta(p_f-p_1) \delta(p_2-p_3)
  \delta(q_f-q_1) \delta(q_2-q_3)
\end{eqnarray}
and the second one
\begin{eqnarray}
\lefteqn{\left< 0 \right| d(q_f) b(p_f) d(q_1) b(p_1) b^+(p_2)
  d^+(q_2) b^+(p_3) d^+(q_3) \left| 0 \right> =}\qquad\nonumber\\
&&\delta(p_1-p_2) \delta(q_1-q_2) \delta(p_f-p_3) \delta(q_f-q_3)
  \nonumber\\
&&  - \delta(p_1-p_2) \delta(p_f-p_3) \delta(q_f-q_2) \delta(q_1-q_3)
  \nonumber\\
&&- \delta(p_f-p_2) \delta(p_1-p_3) \delta(q_1-q_2) \delta(q_f-q_3)
  \nonumber\\
&&  + \delta(p_f-p_2) \delta(p_1-p_3) \delta(q_f-q_2) \delta(q_1-q_3)
{}.
\end{eqnarray}

Terms of the form $\delta(p_2-p_3) \delta(q_2-q_3)$ describe closed
fermion loops, and we drop them in the calculation of the reduced
amplitude.  Therefore only two terms remain.  In third order we get
\begin{equation}
\left< f \right| S^{(3)} \left| i \right> = - i \frac{2}{3!}
  \gamma(p_f) \gamma(q_f) \int d\Gamma(p_1) d\Gamma(q_1)
  V(p_f,q_1) V^*(p_1,q_1) V(p_1,q_f)
{}.
\end{equation}
Combining both results we get up to third order
\begin{equation}
\left< f \right| S^{(1+3)} \left| i \right> = \gamma(p_f)
  \gamma(q_f) \left[ M(p_f,q_f) + \frac{1}{3} \int d\Gamma(p_1)
  d\Gamma(q_1) M(p_f,q_1) M^*(p_1,q_1) M(p_1,q_f) \right]
\label{Magnus to third order}
{}.
\end{equation}

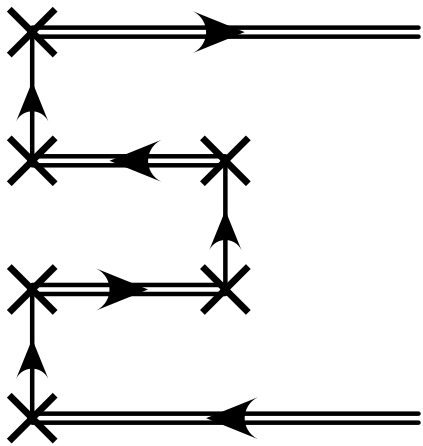
\begin{figure}
\caption{Third order correction to the one-pair creation process in
second-order Magnus theory; compare with
Fig.~\protect\ref{Fig:PViol}.  Double lines mean that the particles
are on shell.}

\label{Fig:higher}
\end{figure}

The interpretation of this higher-order process is straightforward.
If we keep in mind that $M(p,q)$ is the amplitude for the production
of a pair and $M^*(p,q)$ the amplitude for the annihilation of a
pair, we get the Feynman diagram as shown in Fig.~\ref{Fig:higher}.
These are just those higher-order processes, we already discussed
before in Sec.~\ref{Sec:Comparision} (see Fig.~\ref{Fig:PViol}).
Two pairs are created and then one of the electrons annihilates with
the other positron, so that finally only one electron and one
positron remain.  The Magnus theory restricts these intermediate
particles to the mass shell, so that we only need the on-shell $M$,
whereas they may be off-shell in the general case.

Using the same technique higher orders can also be calculated.  The
explicit calculation of the fifth and seventh order gives
\begin{mathletters}
\begin{eqnarray}
\left< f \right| S^{(5)} \left| i \right> &=& \frac{16}{5!}
  \gamma(p_f) \gamma(q_f) \int d\Gamma(p_1) d\Gamma(q_1) d\Gamma(p_2)
  d\Gamma(q_2) \nonumber\\
&&\times M(p_f,q_1) M^*(p_1,q_1) M(p_1,q_2) M^*(p_2,q_2) M(p_2,q_f)\\
\left< f \right| S^{(7)} \left| i \right> &=& \frac{272}{7!}
  \gamma(p_f) \gamma(q_f) \int d\Gamma(p_1) d\Gamma(q_1) d\Gamma(p_2)
  d\Gamma(q_2) d\Gamma(p_3) d\Gamma(q_3) \nonumber\\
&&\times M(p_f,q_1) M^*(p_1,q_1) M(p_1,q_2) M^*(p_2,q_2)
M(p_2,q_3) M^*(p_3,q_3) M(p_3,q_f)
,
\end{eqnarray}
\end{mathletters}
where in the calculation of the vacuum expectation values also more
complicated fermion loops occur, which have to be neglected.  The
following properties are remarkable, as they are true for all higher
orders: First all contributions coming from the same order are of
the same type and can therefore be summed.  Second all amplitudes
have the same sign, that is, they are all added coherently.
Therefore they increase all the reduced total probability, no
cancellation or reduction occurs.  This same sign can also be
understood from the connection of the higher-order processes with
the vacuum corrections.  Vacuum loops contribute with a negative
sign to the amplitude (see Sec.~\ref{Sec:Pertubation} and
\cite{Feynman:QED}).  As our diagrams are connected with a vacuum
diagram by the permutation of some lines, it is clear that it has to
have the same sign as the lowest-order diagram.  The general form of
the $n$th order therefore is

\begin{eqnarray}
\left< f \right| S^{(n)} \left| i \right> &=& c_n \gamma(p_f)
\gamma(q_f) \int d\Gamma(p_1) d\Gamma(q_1) \cdots d\Gamma(p_{(n-1)/2})
d\Gamma(q_{(n-1)/2}) \nonumber\\
&&\times M(p_f,q_1) M^*(p_1,q_1) M(p_1,q_2) \cdots
M^*(p_{(n-1)/2)},q_{(n-1)/2}) M(p_{(n-1)/2},q_f)
,
\end{eqnarray}

For higher orders the number of terms occurring in the reduction of
the vacuum expectation values gets rather large and the number of
diagrams increases rapidly.  However we show in the appendix how
these total number of processes and therefore $c_n$ can be
calculated using combinatorial arguments and a recursive formula.

Finally we calculate the lowest-order correction to the reduced
total probability.  Squaring the pair production amplitude up to
third order (Eq.~(\ref{Magnus to third order})) we get
\begin{eqnarray}
P^{(1+3)}(p_f,q_f) &=&
  \left|\left<f \right| S^{(1+3)}\left|i\right> \right|^2 \nonumber\\
&=& \left[ M^*(p_f,q_f) + \frac{1}{3} \int d\Gamma(p) d\Gamma(q)
  M^*(p_f,q) M(p,q) M^*(p,q_f) \right] \nonumber\\
&&\times \left[ M(p_f,q_f) +
  \frac{1}{3} \int d\Gamma(p') d\Gamma(q') M(p_f,q') M^*(p',q')
  M(p',q_f)\right]\nonumber\\
&&\times \gamma^2(p_f) \gamma^2(q_f)\nonumber\\
&=& \biggl\{|M(p_f,q_f)|^2 + \frac{1}{3} \biggl[ \int d\Gamma(p)
  d\Gamma(q) M(p_f,q_f) M^*(p_f,q) M(p,q) M^*(p,q_f) \nonumber\\
&& + M^*(p_f,q_f) M(p_f,q) M^*(p,q) M(p,q_f)\biggr] + \cdots\biggr\}
  \gamma^2(p_f) \gamma^2(q_f) \nonumber\\
&=& \biggl\{|M(p_f,q_f)|^2 + \frac{2}{3} \int d\Gamma(p) d\Gamma(q)
  \nonumber\\
&&\times  \mbox{Re} \left[M(p_f,q_f) M^*(p_f,q) M(p,q) M^*(p,q_f)
  \right] + \cdots\biggr\} \gamma^2(p_f) \gamma^2(q_f)
{}.
\end{eqnarray}

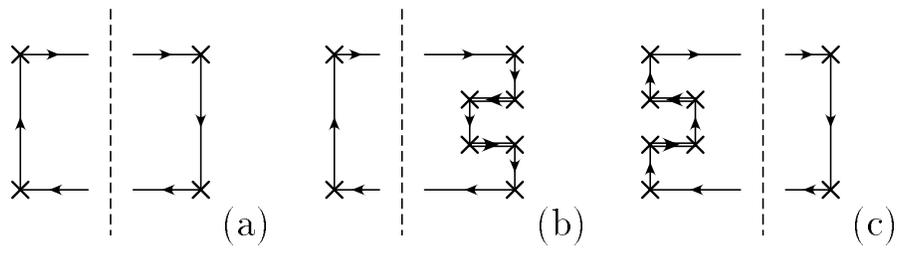
\begin{figure}
\caption{Lowest order (a) and the first two
correction terms (b and c) to the reduced one-pair
creation probability.}
\label{Fig:lowfirst}
\end{figure}

Integrating over $p_f$ and $q_f$ we get the total probability
(Fig.~\ref{Fig:lowfirst})
\begin{eqnarray}
P_{\rm total} &=& \int d\Gamma(p_f) d\Gamma(q_f) |M(p_f,q_f)|^2
  + \frac{2}{3} \int d\Gamma(p_f) d\Gamma(q_f) d\Gamma(p) d\Gamma(q)
\nonumber\\
&&\times \mbox{Re}\left[ M(p_f,q_f) M^*(p_f,q) M(p,q) M^*(p,q_f)
  \right] \\
\label{ZigZag}
&=:& P^{(B)} + P^{(M)}
{}.
\end{eqnarray}

This result is easily generalized. The general series is of the form
\begin{eqnarray}
P_{\rm total} &=& \int d\Gamma(p_1) d\Gamma(q_1) |M(p_1,q_1)|^2
  \nonumber\\
&& + d_3 \int d\Gamma(p_1) d\Gamma(q_1) d\Gamma(p_2) d\Gamma(q_2)
  \mbox{Re}\left[ M(p_1,q_1) M^*(p_1,q_2) M(p_2,q_2) M^*(p_2,q_1)
  \right]
  \nonumber\\
&& + \cdots \nonumber\\
&& + d_{2l-1}\int d\Gamma(p_1) \cdots d\Gamma(q_l) \mbox{Re}
\underbrace{\left[ M(p_1,q_1) M^*(p_1,q_2) \cdots
M(p_l,q_l) M^*(p_l,q_1)\right]}_{l \times}
  \nonumber\\
&& + \cdots
{}.
\label{Mg:GeneralP}
\end{eqnarray}

\narrowtext
The coefficients $d_n$ are again calculated in the appendix.
They can be derived easily from the coefficients $c_n$ of the
amplitudes.

\section{Calculation for impact parameter zero}
\label{BornZero}

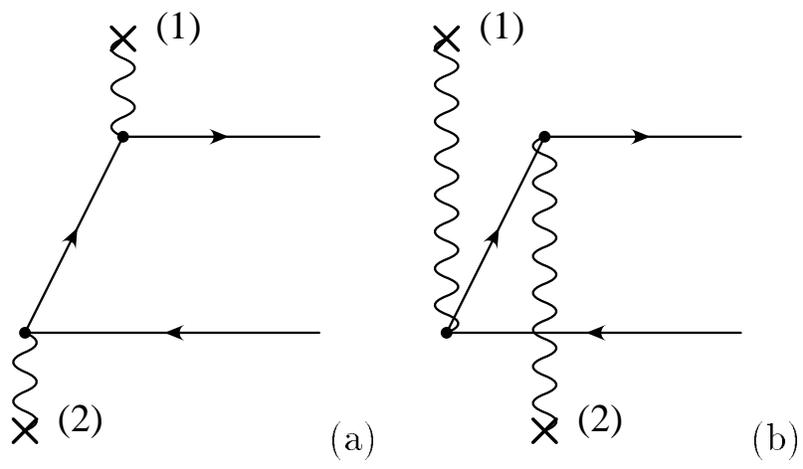
\begin{figure}
\caption{The two Feynman diagrams contributing to the one-pair
creation in lowest order. The interaction with the external field
of ion 1 and 2 is denoted by $(1)$ and $(2)$, respectively.}
\label{Fig:TBorn}
\end{figure}

For the calculation of the lowest-order multiple-particle correction
to the one-pair creation probability we make use of the analytic
form of the matrixelement for impact parameter $b$ zero.  Details of
this calculation can be found in a previous publication
\cite{BornZero}, therefore we will only review rather briefly the
main properties.  The Feynman diagrams contributing to the lowest
order Born calculation are shown in Fig.~\ref{Fig:TBorn}, where (1)
and (2) denotes the interaction with the external field of ion 1 and
2 , respectively.  The matrix element is
\begin{eqnarray}
\lefteqn{M(p,q) = -i e^2 \bar u(p) }\qquad\nonumber\\
&&\times  \biggl[ \int \frac{d^4k}{(2 \pi )^4}
  {\not\!\!A}_1(p - k) \frac{{\not\!k} + m}{k^2 - m^2}
  {\not\!\!A}_2(q + k)  \nonumber\\
&&+ \int \frac{d^4k}{(2 \pi )^4}
  {\not\!\!A}_2(p - k) \frac{{\not\!k} + m}{k^2 - m^2}
  {\not\!\!A}_1(q + k)  \biggr] \;v(q)
{}.
\end{eqnarray}
The external fields are given by
\begin{equation}
A_\mu^{(1,2)} (q) = - 2 \pi Z_{(1,2)} e u_\mu^{(1,2)}
  \delta (q u^{(1,2)}) \frac{F(q^2)}{q^2}
{}.
\end{equation}
$u^{(1,2)}$ are the four velocities of the ions and $F(q^2)$ is
their form factor.  All calculations are done for collisions of the
same type of ions and in the center of mass system.  For the form
factor a simple dipole form factor of the form
\begin{equation}
  F_{\rm dipole}(q^2) = \frac{\Lambda^2}{\Lambda^2-q^2}
,
\end{equation}
with $\Lambda = 83 $MeV, has been used together with a second form
factor, which is a linear combination of two dipole form factors.
In the previous publication the integral has been solved in order to
calculate the lowest-order Born result.  Here we use this analytic
result for $M$ in Eq.~(\ref{ZigZag}).  The usual way to calculate
expressions of this kind is to rewrite the spin summation as a trace
over gamma matrices.  But as each of the $M$ consists of two
diagrams, we finally get a total of sixteen different traces, which
all give large expressions, which are not manageable by an algebraic
calculation program and therefore can not be used for numerical
calculations.  Therefore we prefer to calculate $M$ directly and to
do the spin summation numerically.

Several methods have been proposed in the literature how to
calculate amplitudes directly instead of their squares
\cite{Bjorken:Chen:67,Kleiss:Stirling:85,Yehudai:92}.  We use the
method described by Fearing and Silbar \cite{Fearing:Silbar:72}.  It
consists of multiplying and dividing $M$ with $\bar v(q) u(p)$ in
order to get
\begin{eqnarray}
M &=& \bar u(p) \hat M v(q) \nonumber\\
&&= \frac{1}{\bar v(q) u(p)} \mbox{Tr}\biggl\{
  v(q)\bar v(q) u(p) \bar u(p) \hat M \biggr\} \nonumber\\
&&= \frac{1}{\bar v(q) u(p)} \mbox{Tr} \biggl\{
  \left( \not\!q - m\right) \frac{1}{2}
  \left( 1 + \gamma_5 \lambda_q \not\!s_q \right) \nonumber\\
&&\times  \left( \not\!p + m\right) \frac{1}{2}
  \left( 1 + \gamma_5 \lambda_p \not\!s_p \right)
  \hat M \biggr\} \nonumber\\
&&= \frac{1}{4 S(p,q)} \mbox{Tr} \biggl\{
  \left( \not\!q - m\right) \nonumber\\
&&\times  \left( 1 + \gamma_5 \lambda_q \not\!s_q
  + \gamma_5 \lambda_p \not\!s_p
  - \lambda_p \lambda_q \not\!s_q \not\!s_p \right) \nonumber\\
&&\times  \left( \not\!p + m\right) \hat M \biggr\}
,\ \
\end{eqnarray}
where $s_p$, $s_q$ are the spinvectors of the electron and the
positron, and $\lambda_p$, $\lambda_q$ are the eigenvalues of the
spinors with respect to the spinvector, that is, they have values of
$\pm 1$.  $S(p,q)$ is a complex number, which has been calculated
using an explicit form for the spinors for the standard form of the
gamma matrices.  In order to make the numerical calculations easier
we have used polarisation vectors $s_p$ and $s_q$, which are
longitudinal vectors:
\begin{equation}
s_p = \frac{1}{\sqrt{p_0^2-p_z^2}} (p_z,0,0,p_0)
\end{equation}
and similar for $s_q$.  The calculation of the trace has been done
with the help of the algebraical calculation program {\sc form}
\cite{FORM:Vermaseren}.  All sixteen different spin combinations
have been calculated and summed.  The final expression has to be
real, which has been used to test the program (As the final
expression can be expressed as a trace, which does not contain any
$\gamma_5$ matrices, it must be equivalent to an expression
containing only real numbers and real scalar products, therefore it
has to be real).  The integration over four particles gives a
12-dimensional integral, of which one angular integration is
trivial.  It was calculated using a Monte Carlo (MC) integration
routine ({\sc vegas} \cite{VEGAS:Lepage:78,VEGAS:Lepage:80}).
Together with the total probability we have calculated also single
differential probabilities by sorting the points into appropriate
bins.  The error of the total probability is always less then 1\%.
An explicit error estimate is given for the differential
probabilities.  As our calculation is symmetric with respect to all
four momenta and all four electron-positron combinations, this error
has been calculated from the standard deviation of the four results.

Finally we want to point out that the expression in
Eq.~(\ref{ZigZag}) can be interpreted also in a different way: If we
calculate the reduced total two-pair creation probability exactly in
lowest order according to the previous section, we get
\begin{eqnarray}
\lefteqn{P^{(2)}_{\rm total} = \frac{1}{(2!)^2}
  \int d\Gamma(p_1) d\Gamma(p_2) d\Gamma(q_1) d\Gamma(q_2)
  }\  \nonumber\\
&&\times
  \left[M^*(p_1,q_1) M^*(p_2,q_2) - M^*(p_1,q_2) M^*(p_2,q_1)
  \right]\nonumber\\
&&\times \left[M(p_1,q_1) M(p_2,q_2) - M(p_1,q_2) M(p_2,q_1)
  \right]\nonumber\\
&&= \frac{1}{2}
  \int d\Gamma(p_1) d\Gamma(p_2) d\Gamma(q_1) d\Gamma(q_2)
  \nonumber\\
&&\times
 \biggl\{ \left|M(p_1,q_1)\right|^2 \left|M(p_1,q_1)\right|^2
\nonumber\\
&&  - \mbox{Re} \left[M^*(p_1,q_1) M(p_2,q_1) M^*(p_2,q_2)
  M(p_1,q_2)\right]\biggr\}
  \nonumber\\
&&= P^{(D)}_{\rm total} - P^{(X)}_{\rm total}
{}.
\end{eqnarray}

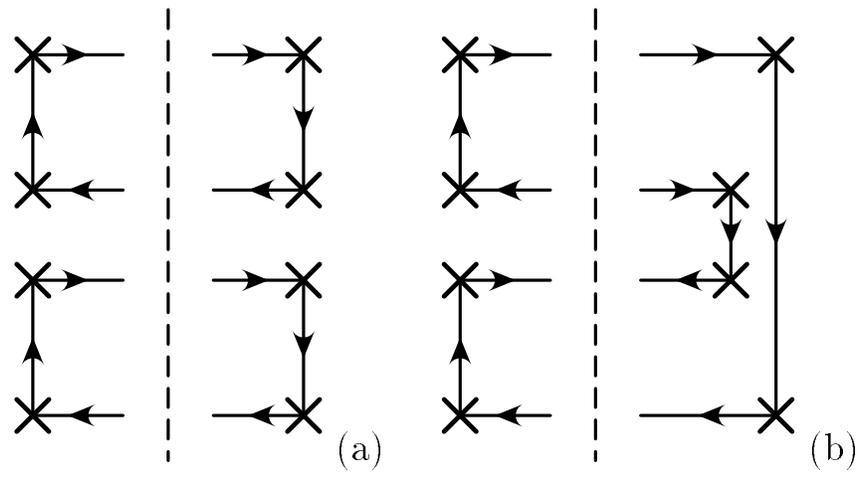
\begin{figure}
\caption{The two Feynman diagrams for the lowest-order two-pair
creation process; direct term (a) and exchange term (b).}
\label{Fig:TwoPair}
\end{figure}

The first term is the direct part, which is used in order to get the
Poisson distribution, whereas the second term is the exchange part,
which we neglect in the Poisson distribution (see also
Fig.~\ref{Fig:TwoPair}).  Comparing the second expression with that
in Eq.~(\ref{ZigZag}) we see that they are identical.  Comparing
Fig.~\ref{Fig:TwoPair} with Fig.~\ref{Fig:lowfirst} we see that we
only have to change the interpretation of the diagram by cutting it
at a different point.  Therefore the result of the calculation for
the total probability can also be used to test, whether the
neglection of the exchange term in order to get the Poisson
distribution is justified.

\section{Results and conclusions}
\label{Results}

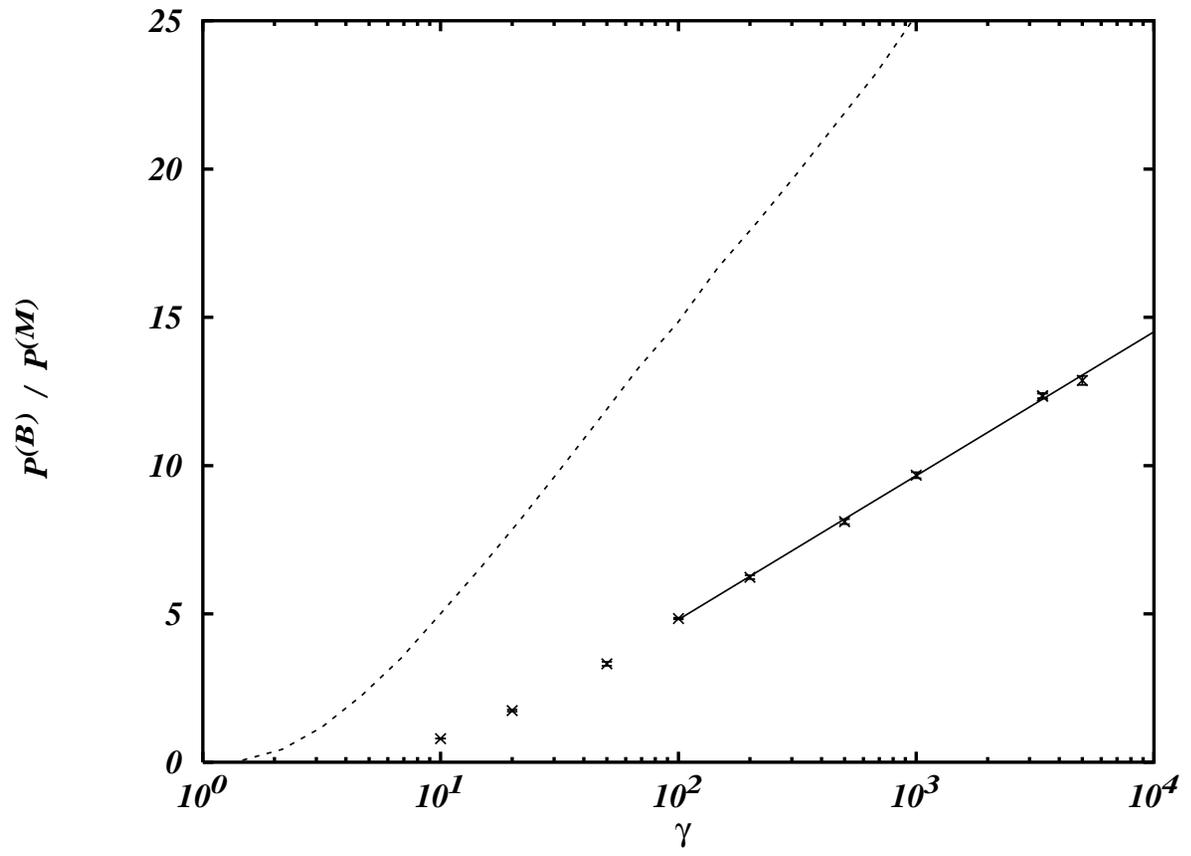
\begin{figure}
\caption{Comparison of the correction to the reduced one-pair
creation probability with the lowest-order result
as a function of $\gamma$. $Z\alpha$ is set to 1.
The points are the results of the calculation of
the multiple-particle corrections, the solid line a fitted
$\ln\gamma$ dependence. The dotted line is the result of the
lowest-order Born approximation.}
\label{Fig:ZZgamma}
\end{figure}

Figure~\ref{Fig:ZZgamma} shows the results of our calculations for
the reduced probability as a function of $\gamma$.  In all our
diagrams we set $Z\alpha = 1$, as this is a common factor in all
results.  Also shown is the result for the corresponding probability
of the lowest-order calculation (see \cite{BornZero}).  In order to
test the dependence on the form factor the calculations have been
done for the dipole form factor and the double dipole form factor.
Both agree within the error interval, therefore we show here only
the results for the double dipole form factor.  We see that the
correction is similar in size as the lowest-order result for large
values of $\gamma$.  Using realistic values for $Z\alpha$ its
importance is reduced, as the higher-order correction has to be
multiplied by $(Z\alpha)^8$, the lowest-order result by
$(Z\alpha)^4$.  In Table~\ref{ZZtable} we give predictions for the
contribution of the higher-order correction to the reduced total
probability.  For $\gamma$ and $Z$ we have used typical values for
relativistic heavy-ion colliders.  The higher-order correction
increases the reduced probability for Pb and U up to about 5--10\%.
Therefore they should be observable in principle.  Based on this
calculation we expect that higher-order corrections (5th order and
more) are again only of about 5--10\% of the third-order results and
are therefore corrections of less than 1\% to the total probability.

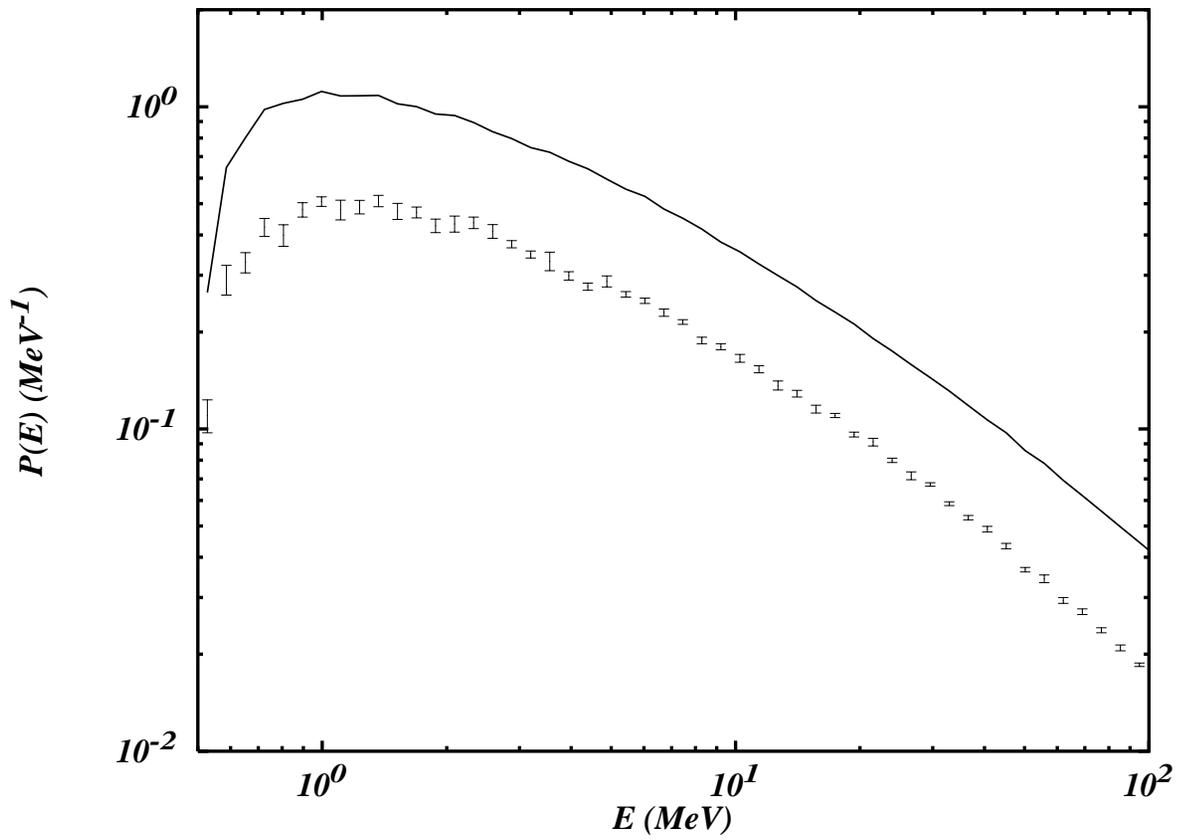
\begin{figure}
\caption{Comparison of the differential probability $P(E)$ for
$\gamma=3400$. Data points are the results of the calculation of
the multiple-particle correction, the solid line results of the
lowest-order Born calculation. $Z\alpha$ is set to 1.}
\label{Fig:ZZPE}
\end{figure}

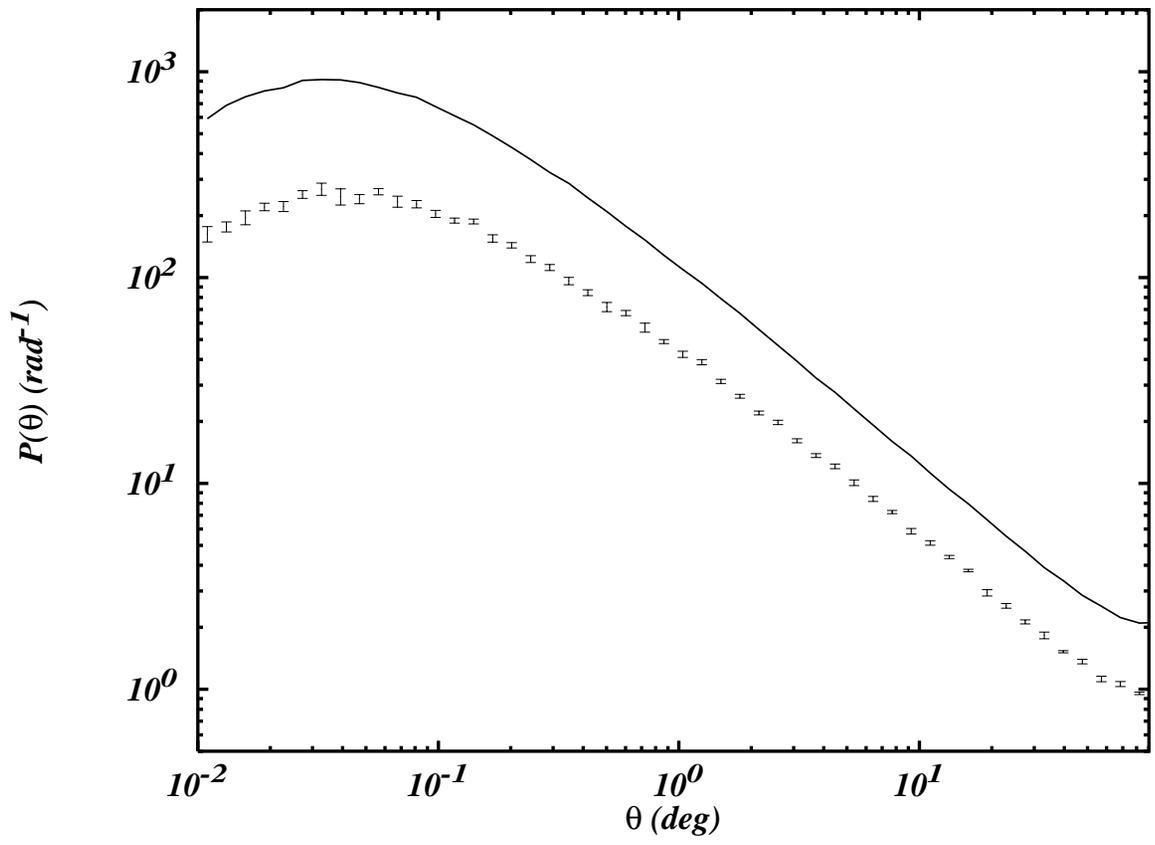
\begin{figure}
\caption{Comparison of the differential probability $P(\theta)$ for
$\gamma=3400$.
Definitions are the same as in Fig.~\protect\ref{Fig:ZZPE}.}
\label{Fig:ZZPtheta}
\end{figure}

Also shown are the results for the differential probabilities as a
function of the energy $E$ (Fig.~\ref{Fig:ZZPE}) and the angle with
the beam axis $\theta$ (Fig.~\ref{Fig:ZZPtheta}).  In these as in
all other single differential probabilities as well the correction
follows more or less the lowest-order result.

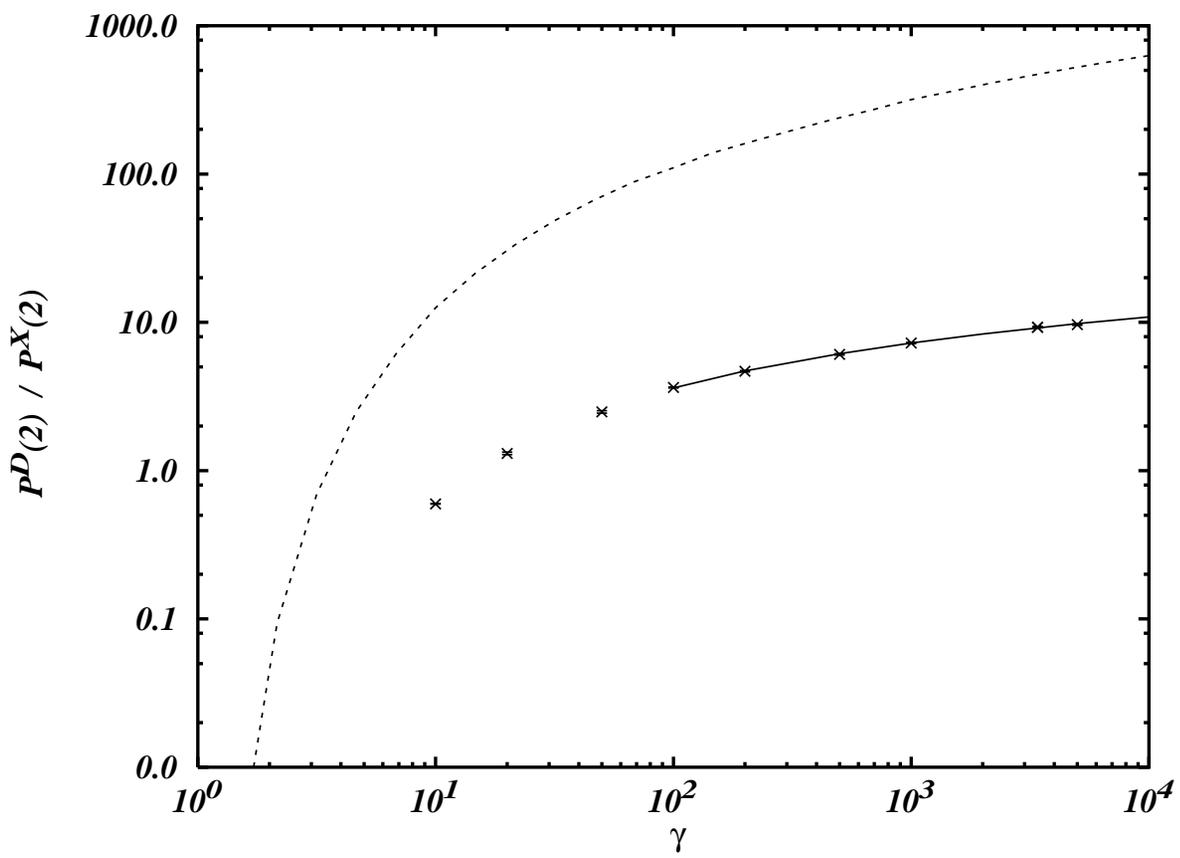
\begin{figure}
\caption{Comparison of the contributions of the direct process
$P^{(D)}$ (dotted line) and the exchange process $P^{(X)}$ (solid
line and data points) to the reduced two-pair creation process.
$Z\alpha$ is set to 1.}
\label{Fig:ZZTwoDX}
\end{figure}

In Fig.~\ref{Fig:ZZTwoDX} we compare both terms contributing to the
total two-pair production probability.  Here the first and the
second diagrams of Fig.~\ref{Fig:TwoPair} are multiplied by the same
factor $(Z\alpha)^8$, therefore the ratio of both curves gives
directly the contribution of the exchange term to the first term,
which is used in the Poisson distribution alone.  We see that the
exchange diagram contributes only with about 1\% to the reduced
total probability.  Also its importance gets smaller for larger
values of $\gamma$, therefore the use of the Poisson distribution
seems to be justified.  This could be different, of course, if one
looks at differential probabilities in some region of the phase
space.

Let us summarize our results: We have shown that the $N$-pair
creation amplitude can be reduced exactly to the reduced one-pair
creation amplitudes and the vacuum amplitude.  Therefore for the
calculation of all $N$-pair probabilities, it suffices to calculate
those two amplitudes.  The use of the external field approximation
and the neglection of the interaction among electrons and positrons
is essential in order to get this result.  We have shown that we get
a Poisson distribution for the total probabilities very generally,
if we neglect all exchange terms.  A calculation of the reduced
one-pair creation probability then suffices alone, as all
higher-order processes are given by the Poisson distribution.

We have found perturbation theory expressions for the reduced
amplitude as well as for the vacuum amplitude, so that both can in
principal be calculated using perturbation theory.  None of them is
of a principal nonpertubative character.

We have compared this result with earlier calculation, where we have
pointed out that they have mainly made use of a quasi boson
approximation of the electron-positron pairs in order to get the
Poisson distribution.  On the other hand our form of the Poisson
distribution needs only the neglection of the exchange terms,
therefore higher-order processes beyond the quasi-boson model can
easily be incorporated.

Based on the general form of the $S$ operator in second-order Magnus
theory we have calculated the lowest-order contribution of such
higher-order multiple-particle effects to the total probabilities
for collisions with impact parameter $b$ zero.  The contribution
were found to be on the 1 -- 10\% level for high $Z$, and should
therefore be incorporated into the calculations.

Finally we compared the deviation of the two-pair creation
probability from the Poisson result.  This deviation was found to be
small, therefore the use of the Poisson distribution seems to be
justified.

\appendix

\section{Coefficients in Magnus theory}
\label{Ap:MgCO}

As shown in Sec.~\ref{Sec:Magnus} all higher-order processes in the
Magnus theory without Coulomb rescattering terms are of the same
type.  The number of terms in each order is given by the reduction
of the vacuum expectation values.  The explicit reduction is not
useful because the complexity of the expressions increases quickly.
Therefore we show here how these can be calculated more easily.

This is a combinatorial problem, which can be formulated in the
following way: In $N$th order we have $N$ interaction points, each
of which either corresponds to a pair creation or pair annihilation
process.  In terms of the fermion lines this means that a line
entering one point from the left also leaves it to the left and the
same from the right.  We need the total number of paths through
these $N$ points using the above condition, where the line initially
enters from the right and finally leaves to the right.  As we look
only at reduced amplitudes, no closed loops are allowed.

Let us find a recursive formula for the number of paths: We define
$A(n,i)$ as the number of distinct paths for $i$ fermion lines
(coming from and leaving to the right) going through $n$ interaction
points.  If we add now another interaction point to the right, there
are two possibilities: If it is a pair creation process, the number
of lines is increased by one.  If we have a pair annihilation
process, we have to connect it with one outgoing and one ingoing
line, therefore reducing their number by one.  For this we have to
choose one of the $i$ outgoing lines and one of the ingoing lines,
where we have to be careful not to make a closed loop; therefore
only $i-1$ of them are allowed.  This means that there are $i(i-1)$
possible combinations.  This gives us the recursion relation
\begin{equation}
  A(n+1,i) = A(n,i-1) + i (i+1) A(n,i+1)
,
\end{equation}
together with the boundary conditions
\begin{mathletters}
\begin{eqnarray}
  A(1,1) & = & 1 \\
  A(1,i) & = & 0 \qquad \mbox{for $i>1$} \\
  A(n,0) & = & 0 \qquad \mbox{for all $n$}
{}.
\end{eqnarray}
\end{mathletters}

With this recursion relation we can calculate $A(n,1)$ for all $n$.
This is just the number of diagrams that we were looking for.
Dividing it by $n!$ gives the coefficient $c_n$ in the expansion of
the amplitude.  These coefficients are given in Table~\ref{MC:cn}.
For the coefficients in the reduced total probability we have to
multiply $c_l$ and $c_{n+1-l}$ and sum over all $l \in 0 \ldots n$.
These are the coefficients $d_n$ of Eq.~(\ref{Mg:GeneralP}).  They
are also given in Table~\ref{MC:cn}.

\narrowtext
\begin{table}
\begin{tabular}{lrlddd}
&$\gamma$&Ion&$P^{(B)}$&$P^{(M)}$& \\
\hline
SPS  & 10   & Pb & 0.63 & 0.013 & 2.1 \% \\
RHIC & 100  & Au & 1.6  & 0.059 & 3.7 \% \\
LHC  & 3400 & Pb & 3.9  & 0.21  & 5.3 \% \\
LHC  & 3400 & U  & 6.1  & 0.49  & 8.1 \% \\
\end{tabular}
\caption{Comparison of the contribution of the
multiple-particle correction $P^{(M)}$ to the
reduced probability with the lowest-order Born result $P^{(B)}$
(see Eq.~(\protect\ref{ZigZag})).}
\label{ZZtable}
\end{table}

\mediumtext
\begin{table}
\begin{tabular}{rrdd}
$n$& \# Diags. & $c_n$ & $d_n$\\
\hline
 1 & $ 1.0000\times 10^{0} $&  1.0000000000 &  1.0000000000\\
 3 & $ 2.0000\times 10^{0} $&  0.3333333333 &  0.6666666667\\
 5 & $ 1.6000\times 10^{1} $&  0.1333333333 &  0.3777777778\\
 7 & $ 2.7200\times 10^{2} $&  0.0539682540 &  0.1968253968\\
 9 & $ 7.9360\times 10^{3} $&  0.0218694885 &  0.0974955908\\
11 & $ 3.5379\times 10^{5} $&  0.0088632355 &  0.0466976645\\
13 & $ 2.2368\times 10^{7} $&  0.0035921280 &  0.0218375158\\
15 & $ 1.9038\times 10^{9} $&  0.0014558344 &  0.0100304665\\
\end{tabular}
\caption{The number of diagrams and the coefficients $c_n$ and
$d_n$ appearing in Magnus theory for different orders $n$.}
\label{MC:cn}
\end{table}


\begin{references}
\bibitem{Baur:Bertulani:88} C.A.~Bertulani and G.~Baur,
Phys.~Rep. {\bf 163}, 299 (1988).

\bibitem{Baur:Unitarity:90} G. Baur,
Phys. Rev. {\bf A42}, 5736 (1990).

\bibitem{Baur:letter:90} G.~Baur,
Phys.~Rev. {\bf D41}, 3535 (1990).

\bibitem{Rhoades-Brown:91} M.J. Rhoades-Brown and J. Weneser,
Phys. Rev. {\bf A44}, 330 (1991).

\bibitem{Best:92} C. Best, W. Greiner, and G. Soff,
Phys. Rev. {\bf A46}, 261 (1992).

\bibitem{Ionescu:93} D.C.~Ionescu,
Phys. Rev. A{\bf 49}, 3188 (1994).

\bibitem{BornZero} K. Hencken, D. Trautmann, and G. Baur,
Phys. Rev. A{\bf 49}, 1584 (1994).

\bibitem{Feynman:49} R. P. Feynman,
Phys. Rev. {\bf 76}, 749 (1949).

\bibitem{Schwinger:54} J. Schwinger,
Phys. Rev. {\bf 93}, 615 (1954).

\bibitem{Itzykson:Zuber:80} C.~Itzykson and J.-B.~Zuber,
{\em Quantum Field Theory} (McGraw-Hill, 1980).

\bibitem{Bialynicki-Birula:75} I. Bia\l ynicki-Birula
and Z. Bia\l ynicki-Birula,
{\em Quantum Electrodynamics} (Pergamon, 1975).

\bibitem{Schweber:61} S.S.~Schweber,
{\em An Introduction to Relativistic Quantum Field Theory}
(Row, Peterson, 1961).

\bibitem{Lippert:91} T. Lippert, J. Thiel, N. Gr\"un, and
W. Scheid,
Int. J. Mod. Phys. {\bf 6A}, 5249 (1991).

\bibitem{Feynman:QED} R.P. Feynman,
{\em Quantum Electrodynamics} (W.A.~Benjamin, 1961).

\bibitem{Greiner:85} W.~Greiner, B.~M\"uller, and J.~Rafelski,
{\em Quantum Electrodynamics of Strong Fields} (Springer, 1985).

\bibitem{Bogoliubov:Shirkov:83} N.N.~Bogoliubov and D.V.~Shirkov,
{\em Quantum Fields} (Benjamin/Cummings, 1983).

\bibitem{Bialynicki-Birula:69} I. Bia\l ynicki-Birula,
B. Mielnik, and J. Pleba\'nski,
Ann. Phys. (NY) {\bf 51}, 187 (1969).

\bibitem{Pechukas:Light:66} P. Pechukas and J.C. Light,
J. Chem. Phys. {\bf 44}, 3897 (1966).

\bibitem{Bjorken:Drell:66} J.D.~Bjorken and S.D.~Drell,
{\em Relativistische Quantenmechanik} (BI Bd. 98, 1966).

\bibitem{Bjorken:Drell:67} J.D.~Bjorken and S.D.~Drell,
{\em Relativistische Quantenfeldtheorie} (BI Bd. 101, 1967).

\bibitem{Jauch:Rohrlich:76} J.M. Jauch and F. Rohrlich,
{\em The Theory of Photons and Electrons} (Springer, 1976).

\bibitem{Bjorken:Chen:67} J.D. Bjorken and M.C. Chen,
Phys. Rev. {\bf 154}, 1335 (1967).

\bibitem{Kleiss:Stirling:85} R. Kleiss and W.J. Stirling,
Nucl. Phys. {\bf B262}, 235 (1985).

\bibitem{Yehudai:92} E. Yehudai, Stanford Linear Accelerator Report
No. SLAC-PUB-92/256-T 1992 (submitted to Phys. Rev. D).

\bibitem{Fearing:Silbar:72} H. W. Fearing and R. R. Silbar,
Phys. Rev. {\bf D6}, 471 (1972).

\bibitem{FORM:Vermaseren} {\sc form} is an algebraical calculation
program by J.A.M. Vermaseren, the free version 1.0 can be found, e.g.,
at {\sc ftp.nikhef.nl}.

\bibitem{VEGAS:Lepage:78} G.P. Lepage,
J. Comp. Phys. {\bf 27}, 192 (1978).

\bibitem{VEGAS:Lepage:80} G.P. Lepage,
Cornell laboratory for nuclear sciences report No.
CLNS-80/447, 1980 (unpublished).
\end{references}
\end{document}